\documentclass[preprint,floatfix,preprintnumbers,amsmath,amssymb,aps]{revtex4}

\usepackage[english]{babel}
\usepackage{graphicx}
\usepackage{dcolumn}
\usepackage{bm}
\usepackage[usenames,dvipsnames]{xcolor}
\usepackage[colorlinks=true,linkcolor=blue,citecolor=blue,urlcolor=blue]{hyperref}
\usepackage{siunitx}
\usepackage{makecell}
\usepackage{multirow}
\usepackage{dsfont}
\usepackage{soul}
\usepackage{braket}
\usepackage{upgreek}

\newcommand{\MC}[1]{\mathcal{#1}}
\newcommand{\MR}[1]{\mathrm{#1}}
\newcommand{\ud}{\mathrm{d}}
\newcommand{\iu}{\mathrm{i}}

\newcommand{\VEC}[1]{\mathbf{#1}}

\newcommand{\MUB}{\mu_{\text{B}}}

\begin{document}

\title{Engineering elliptical spin-excitations by complex anisotropy fields in Fe adatoms and dimers on Cu(111)}
\author{Filipe S. M. Guimar\~aes}
\email{f.guimaraes@fz-juelich.de}
\author{Manuel dos Santos Dias}
\author{Benedikt Schweflinghaus}
\author{Samir Lounis}
\email{s.lounis@fz-juelich.de}
\affiliation{Peter Gr\"unberg Institut and Institute for Advanced Simulation, Forschungszentrum J\"ulich \& JARA, D-52425 J\"ulich, Germany}

\date{\today}

\begin{abstract}

We investigate the dynamics of Fe adatoms and dimers deposited on the Cu(111) metallic surface in the presence of spin-orbit coupling, within time-dependent density functional theory. 
The \textit{ab initio} results provide material-dependent parameters that can be used in semiclassical approaches, which are used for insightful interpretations of the excitation modes.
By manipulating the surroundings of the magnetic elements, we show that elliptical precessional motion may be induced through the modification of the magnetic anisotropy energy.
We also demonstrate how different kinds of spin precession are realized, considering the symmetry of the magnetic anisotropy energy, the ferro- or antiferromagnetic nature of the exchange coupling between the impurities, and the strength of the magnetic damping. 
In particular, the normal modes of a dimer depend on the initial magnetic configuration, changing drastically by going from a ferromagnetic metastable state to the antiferromagnetic ground state.
By taking into account the effect of the damping into their resonant frequencies, we reveal that an important contribution arises for strongly biaxial systems and specially for the antiferromagnetic dimers with large exchange couplings.
Counter intuitively, our results indicate that the magnetic damping influences the quantum fluctuations by decreasing the zero-point energy of the system.

\end{abstract}

\maketitle

\section{Introduction}

%
%

Future technological devices demand an understanding of quantum mechanisms in nanostructures such as single atoms and small clusters \cite{Wiesendanger:2009hg,Ternes:2017gm}.
Recent atomic manipulation and spectroscopy experiments utilizing the scanning tunneling microscope (STM) pushed forward the frontiers of this area by the development of logic operations based on atomic spin manipulation \cite{Khajetoorians:2011gx}, subnanometer-sized sensors \cite{Choi:2017hg}, magnetic stability of single adatoms \cite{Donati:2016ko}, and many other atomic-scale realizations \cite{Bryant:2013dr,Zhou:2010js,vanBree:2017cm,Natterer:2017ex,IbanezAzpiroz:2017dv,Hirjibehedin:2007eg}.
Additionally to miniaturizing components, advanced spintronic devices also require ultrafast manipulation of the magnetic units. 
With that aim, it is natural to investigate the dynamic processes of those magnetic building blocks \cite{Heinrich:2004dh,Hirjibehedin:2007eg,Schuh:2012ih,dosSantosDias:2015bh,Khajetoorians:2010bh,Chilian:2011bt,Khajetoorians:2013hj,Spinelli:2014db,Jacobson:2015kk,Dubout:2015eb}.

Many of those atomic-scale investigations are made on insulating surfaces, where the host weakly interacts with the magnetic units (see, e.g., Refs. \onlinecite{Hirjibehedin:2007eg,FernandezRossier:2009fs,Persson:2009gd,Fransson:2009hs,Sothmann:2010ci,Bryant:2013dr,Spinelli:2014db,Donati:2016ko}).
On the other hand, for metallic hosts, the surface and deposited structures hybridize strongly, modifying the original electronic states of the isolated subsystems \cite{Lounis2010,Khajetoorians:2010bh,Chilian:2011bt,Khajetoorians:2013hj,Khajetoorians:2013gq}.
This strong coupling leads to a noninteger magnetic moment and to broadened spin excitation spectra of the deposited nanostructures.
It also affects the magnetic excitations of those systems, changing relaxation times \cite{Gilbert:2004gx} and influencing the polarization of spin currents pumped out of the magnetic unit into the substrate \cite{Tserkovnyak:2002ju}.
These spin currents may be used to excite other magnetic units deposited on the surface \cite{Guimaraes:2010fi}, or, in the presence of spin-orbit coupling, to generate charge currents \cite{Ando:2011dg,Jiao:2013kv}.

When two or more magnetic atoms or clusters are brought together, their mutual interaction can lead to ferromagnetic, collinear antiferromagnetic, or even more complex magnetic structures.
Their excitation modes depend on the local environment and also on the exchange coupling between the components. 
In particular, antiferromagnetic systems have been studied for decades \cite{Anderson:1951fx,Anderson:1952jp,Keffer:1952jw} and still generate interest in its ground state description and possible excitations \cite{Holzberger:2013du,Sandratskii:2012cu}.
These structures play a central role in the rising and promising field of antiferromagnetic spintronics \cite{Jungwirth:2016dt}, in which excitations and switching can be induced by spin-orbit torques, when charge currents are applied to adjacent heavy metal layers \cite{Cheng:2016gq,Guimaraes:2015fl}.

In this paper, we investigate magnetic excitations of the smallest possible nanostructures --- adatoms and dimers, primer constituents of any ferromagnetic or antiferromagnetic system --- deposited on metallic surfaces.
To correctly capture the mixing effects discussed above, we employ first principles calculations based on time-dependent density functional theory \cite{Gross:1985gx,Lounis2010,Lounis:2011bn} taking the effects of the spin-orbit coupling into account, which can lead to nontrivial magnetic anisotropy fields.
We focus on the dynamic transverse magnetic susceptibility that describes the density of spin excitations and is directly related to the measured conductance in inelastic scanning tunneling spectroscopy (ISTS) experiments \cite{Schweflinghaus:2014eq}.
This quantity is defined as the magnetic response of the system to an oscillatory transverse magnetic field, as in ferromagnetic resonance experiments.
We also make use of a semiclassical model to interpret the features of the precessional modes and to define effective parameters.

We chose a prototypical system composed by Fe adatoms and dimers deposited on the Cu(111) surface, but our results apply more generally to transition metals deposited on other metallic surfaces, in a qualitative way.
For a single magnetic adatom, we show how external magnetic fields can stabilize different magnetization directions and drastically change the precession shape of the motion.
We also demonstrate how similar effects can be obtained by atomic engineering \cite{Loth:2010gq,Oberg:2013jv,Bryant:2013dr}, namely by bringing nonmagnetic Cu atoms to the vicinity of the magnetic Fe atom, and thus modifying the magnetic anisotropy energy landscape of the system.
Tilted anisotropies are also important to break the symmetries of spin-orbit torques, leading to deterministic switching of the magnetic unit \cite{You:2015cz}.
When a second magnetic atom is placed close to the first one, the rotational symmetry is naturally broken and a biaxial anisotropy is induced. 
We also take advantage of the distance-dependent oscillatory exchange interaction \cite{Zhou:2010js} to design structures with ferromagnetic (FM) and antiferromagnetic (AFM) alignments, showing how the spin excitations change by varying just the starting magnetic configuration.
For such nanosized magnetic structures, zero-point spin fluctuations can be of concern, and so we shall also discuss their role.

This paper is organized as follows. 
In Sec.~\ref{sec:theory}, we describe the first principles approach using time-dependent density functional theory (TDDFT) that we use to obtain the excitation spectra of the magnetic structures.
We also derive a phenomenological formalism to aid in the analysis of the \textit{ab initio} results.
By fitting the results of the first-principles calculations to this model, we obtain material-dependent parameters that may be used in semiclassical approaches.
Section~\ref{sec:adatom} is devoted to the single adatom excitations. 
The magnetization dynamics of $3d$ adatoms on the Cu(111) surface were explored in Ref.~\onlinecite{dosSantosDias:2015bh}.
Here we expand this study and explore how the anisotropies affect the magnetization dynamics, focusing on how they lead to elliptical precession. 
In Sec.~\ref{sec:dimer}, we investigate the excitations of Fe dimers, with large and relatively small interatomic magnetic couplings. 
In the latter case, we analyze the excitations starting from two different states: when the magnetic moments are aligned ferromagnetically (metastable state) and antiferromagnetically (ground state).
We show how a simple change of the starting alignment drastically alters the excitation spectra depending on the coupling, anisotropy, and damping.
Finally, in Sec.~\ref{sec:Summary} we summarize our results.


\section{Theoretical Framework}
\label{sec:theory}

\subsection{Spin excitations from first-principles}
\label{sec:tddft}

Our first-principles description of spin excitations is based on time-dependent density functional theory, using the linear response approach.
The central object is the dynamical magnetic susceptibility, which is closely related to the inelastic tunneling conductance measured experimentally, as explained in Ref.~\onlinecite{Schweflinghaus:2014eq}.
This approach requires two steps: First the self-consistent ground state electronic structure must be found, and then the linear response of the ground state to a magnetic perturbation is evaluated.
These steps are briefly outlined below.

The ground state electronic structure is obtained from DFT calculations, based on the Korringa-Kohn-Rostoker Green function method~\cite{Papanikolaou:2002hk} (KKR-GF), in the local spin density approximation (LSDA), as parametrized by Vosko, Wilk, and Nusair~\cite{Vosko:1980fd}.
The scattering problem is solved in the atomic sphere approximation (ASA) with $\ell_{\MR{max}}=3$ cutoff, with subsequent use of the full charge density.
Energy integrations are performed with a rectangular contour in the upper complex energy plane using 40 points, including five Matsubara frequencies with temperature $T = \SI{50}{\kelvin}$ \cite{Wildberger:1995gp}.

The electronic structure of the clusters on the Cu(111) surface is calculated in two steps.
First the pristine surface is simulated using a 22-layer slab of Cu(111) planes, augmented with two vacuum regions with the thickness of four bulk layers.
The in-plane lattice constant is the experimental one, $a = 3.615/\sqrt{2}\,\si{\angstrom} = \SI{2.556}{\angstrom}$.
No relaxation of the interlayer distance has been considered for the slab calculation.
A k mesh with $180\times180$ points in the whole two-dimensional Brillouin zone is employed.
Next, a real-space cluster is embedded at the surface, including the adatoms and all surrounding nearest-neighbor positions (Cu atoms and vacuum), and treated self-consistently.
Structural optimization of the supported nanostructures is accounted for by a vertical relaxation of the structure towards the surface by 14\% of the bulk interlayer distance~\cite{Stepanyuk:2003kc,Negulyaev:2008eb}.

The key quantity for the description of spin excitations is the dynamical magnetic susceptibility.
In linear response, when an external monochromatic magnetic field perturbs the system, this susceptibility describes the linear change to the spin density that it causes
\begin{equation}
 \delta M_{\mu}(\VEC{r},\omega) = \sum_\nu \int\!\ud\VEC{r}'\,\chi_{\mu\nu}(\VEC{r},\VEC{r}',\omega)\,\delta B_\nu(\VEC{r}',\omega) \quad.
\end{equation}
Here $\mu,\nu=x,y,z$ are the cartesian components of the spin density.
Within TDDFT, the magnetic susceptibility is related to the one of the Kohn-Sham electrons through the Hartree-exchange-correlation kernel via a Dyson-like equation:
\begin{equation}
 \chi(\omega) = \chi^{\MR{KS}}(\omega) + \chi^{\MR{KS}}(\omega)\,K^{\MR{Hxc}}\,\chi(\omega) \quad,
\end{equation}
where the spatial dependence has been omitted.
The kernel $K^\text{Hxc}$ includes the Hartree and the exchange-correlation contributions.
The adiabatic LSDA was implied, which leads to a frequency-independent kernel.
Further details on the formalism and its implementation can be found in Refs.~\onlinecite{Lounis2010,Lounis:2011bn,Lounis:2014in,dosSantosDias:2015bh,Lounis:2015ho}.

Spin-orbit coupling leads to new aspects in the calculation of the dynamical magnetic susceptibility.
Let $\VEC{M}(\VEC{r})$ be the ground state vector spin density.
Then one may define a pointwise transformation of the global cartesian axes such that in this new, so-called local spin frame of reference, the ground state spin density has only one component,
\begin{equation}\label{frame}
 M(\VEC{r})\,\VEC{\hat{n}}_z = \MC{R}(\VEC{r})\,\VEC{M}(\VEC{r}) \quad.
\end{equation}
Throughout this paper, vector components in the global frame $\{\VEC{\hat{n}}_{\alpha'}\}$ are primed, while in the local frame $\{\VEC{\hat{n}}_{\alpha}\}$ they are unprimed ($\alpha = x, y, z$).
In the local frame, the Dyson-like equation for the dynamic susceptibility has the following matrix structure:
\begin{equation}
 \renewcommand{\arraystretch}{1.2}
 \begin{pmatrix}
  \chi_{\MR{TT}} & \chi_{\MR{TL}} \\
  \chi_{\MR{LT}} & \chi_{\MR{LL}}
 \end{pmatrix}^{\!-1}
\!\!\!=\,\, \begin{pmatrix}
  \chi_{\MR{TT}}^{\MR{KS}} & \chi_{\MR{TL}}^{\MR{KS}} \\
  \chi_{\MR{LT}}^{\MR{KS}} & \chi_{\MR{LL}}^{\MR{KS}}
 \end{pmatrix}^{\!-1}
\!\!\!-\,\, \begin{pmatrix}
  K_{\MR{T}}^{\MR{xc}} & 0 \\
  0 & K_{\MR{L}}^{\MR{Hxc}}
 \end{pmatrix}
 \quad,
\end{equation}
where the $2\times2$ blocks correspond to $\MR{T} = \{x,y\}$ and $\MR{L} = \{z,n\}$, and the index $n$ pertains to the charge density.
Frequency and spatial dependence were omitted for clarity.
Note that the Hartree contribution only appears in the longitudinal part of the kernel, as indicated by the superscript `Hxc'.
The off-diagonal blocks of the KS susceptibility, $\chi_{\MR{LT}}^{\MR{KS}}$ and $\chi_{\MR{TL}}^{\MR{KS}}$, arise due to spin-orbit coupling and/or to noncollinear magnetic structures.
If both of these are weak or absent, we can restrict the calculation only to the purely transverse part of the dynamical susceptibility, $\chi_{\MR{TT}}$, provided the transformation to the local frame has been performed.
This is the situation encountered for the systems described in this work.

The dynamical magnetic susceptibility is a very complex object, containing the information about all kinds of spin excitations as well as their detailed spatial and frequency dependence.
A more intuitive physical picture is afforded by defining atomic like quantities, by integrating out the spatial dependence over a region of space assigned to a given magnetic atom $i$ ($j$):
\begin{equation}
 \chi_{i\mu,j\nu}(\omega) = \int_{V_i}\!\!\!\ud\VEC{r}\int_{V_j}\!\!\ud\VEC{r}'\,\chi_{\mu\nu}(\VEC{r},\VEC{r}',\omega) \quad.
\end{equation}
If the external perturbing magnetic field is taken to be uniform within each atomic volume $V_i$, we arrive at an effective atomic description:
\begin{equation}\label{deltam}
 \delta M_{i\mu}(\omega) = \sum_{j\nu}\chi_{i\mu,j\nu}(\omega)\,\delta B_{j\nu}^{\MR{ext}}(\omega) \quad.
\end{equation}
In the low frequency regime, we can hope to make a connection to atomistic spin dynamics, as explained in the following.

\subsection{Phenomenological approach}
\label{sec:llg}

To interpret the results obtained using the TDDFT formalism described above, we also employ a phenomenological description of the magnetization dynamics given by the Landau-Lifshitz-Gilbert (LLG) equation \cite{Gilbert:2004gx}.
The $2\times2$ transverse magnetic susceptibility $\chi_{\MR{TT}}$ can be obtained by assuming small deviations around the local equilibrium direction, which defines the $\VEC{\hat{n}}_z$ axis.
Using this approach, we also extract all the relevant parameters directly from the first-principles dynamical susceptibility \cite{dosSantosDias:2015bh}.

The equation of motion for the magnetic moment of atom $i$ is given by
\begin{equation}\label{llg}
\begin{split}
\frac{\ud \mathbf{M}_i}{\ud t} = -\gamma\mathbf{M}_i\times\mathbf{B}^\text{eff} + \frac{\alpha}{M_i} \mathbf{M}_i\times \frac{\ud\mathbf{M}_i}{\ud t}
\end{split}\quad.
\end{equation}
The first term on the right-hand side represents the torque due to an effective field $\mathbf{B}^\text{eff}_i =-\partial E/\partial \mathbf{M}_i$ obtained from the energy functional $E(\{\mathbf{M}_i\})$ of the system.
The last term describes relaxation effects that push the magnetization back to the equilibrium orientation.
$\gamma$ is the gyromagnetic ratio, and the damping is characterized by the Gilbert parameter $\alpha$. 
The latter is, in principle, a $3\times3$ matrix that captures its possible anisotropic behavior \cite{Bhattacharjee:2012if}. 
Nevertheless, for the purpose of the discussion set forth in this work, a scalar quantity is sufficient.

Considering an energy functional that includes magnetic anisotropies, coupling between the different components, and an external magnetic field $\mathbf{B}^\text{ext}$ we can write
\begin{equation}\label{energy}
\begin{split}
E(\{\mathbf{M}_i\}) =& \sum_i E_i(\mathbf{M}_i) - \frac{J}{M^2}\,\mathbf{M}_1\cdot\mathbf{M}_2
\end{split}\quad.
\end{equation}
where $i=1,2$ labels the magnetic atoms, and
\begin{equation}\label{singleenergy}
\begin{split}
E_i(\mathbf{M}_i) =& \mathbf{M}^\text{T}_i\frac{\mathbf{K}_i}{M_i^2}\mathbf{M}_i -\mathbf{B}^\text{ext}\cdot\mathbf{M}_i
\end{split}
\end{equation}
is the single atom energy containing the local anisotropy and Zeeman energies.
For the systems we investigate, the general $3\times3$ matrix describing the on-site anisotropy can always be brought to the diagonal form
\begin{equation}\label{kmatrix}
\begin{split}
\mathbf{K}_i = \left(\begin{array}{ccc} K_{ix} & 0 & 0 \\ 0 & K_{iy} & 0 \\ 0 & 0 & 0 \end{array}\right)
\end{split}
\end{equation}
by a suitable definition of the local frame of reference, as explained in connection with Eq.~\eqref{frame}.

When $\mathbf{M}_i$ is in equilibrium, i.e., pointing along the $\VEC{\hat{n}}_z$ direction of the local frame of reference, the effective magnetic field is given by
\begin{equation}\label{beff}
\begin{split}
\mathbf{B}^\text{eff}_i = \left(\frac{J}{M}+ B^\text{ext}\right) \VEC{\hat{n}}_z
\end{split}\quad,
\end{equation}
where we assume that the magnetization is aligned with the external field. 

The spin excitations can be described by the transverse dynamical magnetic susceptibility.
This quantity can be formally derived by calculating the small oscillations of the magnetization $\mathbf{M}_i(t) = M_i\VEC{\hat{n}}_z+\delta\mathbf{M}_i(t)$ induced by a transverse oscillatory external field $\delta\mathbf{B}^\text{ext}(t) = \delta B^\text{ext}_0\left[\cos(\omega t)\VEC{\hat{n}}_x + \sin(\omega t)\VEC{\hat{n}}_y\right]$.
In the frequency domain, the change in the magnetization is given by Eq.~\eqref{deltam}.
We can then map the results obtained from the first-principles calculations to the analytical phenomenological expressions (listed in Appendix~\ref{apx:eqsmotion}). 
The anisotropy constants, exchange coupling, gyromagnetic ratio, and Gilbert damping for each case is obtained by fitting the appropriate functional form to the components of the transverse susceptibility $\chi_{\MR{TT}}$ calculated using TDDFT close to $\omega=0$ \cite{dosSantosDias:2015bh}.

It is convenient to work with the local circular basis $\VEC{\hat{n}}_\pm = \VEC{\hat{n}}_x\pm \iu \VEC{\hat{n}}_y$. 
Using general complex components containing information on the amplitude and phase of the oscillation, the transverse vectors are then transformed from cartesian $(v_x,v_y)$ to circular $(v_-,v_+)$ as $v_\pm = v_x \pm \iu v_y$, and
\begin{equation}\label{changebasis}
\begin{split}
\mathbf{v}(t)=& \operatorname{Re}\left[(v_x\VEC{\hat{n}}_x+ v_y\VEC{\hat{n}}_y) e^{-\iu\omega t}\right]\\
=&\frac{1}{2} \operatorname{Re}\left[(v_-\VEC{\hat{n}}_++ v_+\VEC{\hat{n}}_-) e^{-\iu\omega t}\right]
\end{split}\quad,
\end{equation}
where $\mathbf{v}(t)$ represents any of the following: the external perturbing field $\delta\mathbf{B}^\text{ext}(t)$, the transverse magnetization components $\delta\mathbf{M}_i(t)$, or the effective field $\delta \mathbf{B}^\text{eff}_i(t)$.
Equation~\eqref{llg} can then be written in the form
\begin{equation}\label{suscmatrix}
\begin{split}
\left(\begin{array}{c} \delta M_{i-} \\\delta M_{i+}\end{array}\right) = \sum_j\left(\begin{array}{cc} \chi_{i-,j+} & \chi_{i-,j-} \\ \chi_{i+,j+} & \chi_{i+,j-} \end{array}\right) \left(\begin{array}{c} \delta B^{\MR{ext}}_{j-} \\\delta B^{\MR{ext}}_{j+}\end{array}\right)
\end{split}\quad.
\end{equation}
A counterclockwise circularly polarized excitation field is described by the cartesian components $\delta B_x^\text{ext}=\delta B$ and $\delta B_y^\text{ext} = \iu\delta B$, where $\delta B$ is a real value that describes the amplitude of oscillation. In the circular basis, this field is described by $\delta B_-^\text{ext}=2\delta B$ and $\delta B_+^\text{ext}=0$.

\subsection{Precessional motion}
\label{sec:ellipse}

%

Spin excitations of small magnetic nanostructures can have a complex precessional nature, the most general form being elliptical.
In order to gain useful insights, here we explain how the excitations can be described using only three parameters: the amplitude $A$, the eccentricity $e$, and the tilt angle $\phi$, which are all dynamical.

\begin{figure}[!htb]
 \centering
  \includegraphics[width=0.5\columnwidth]{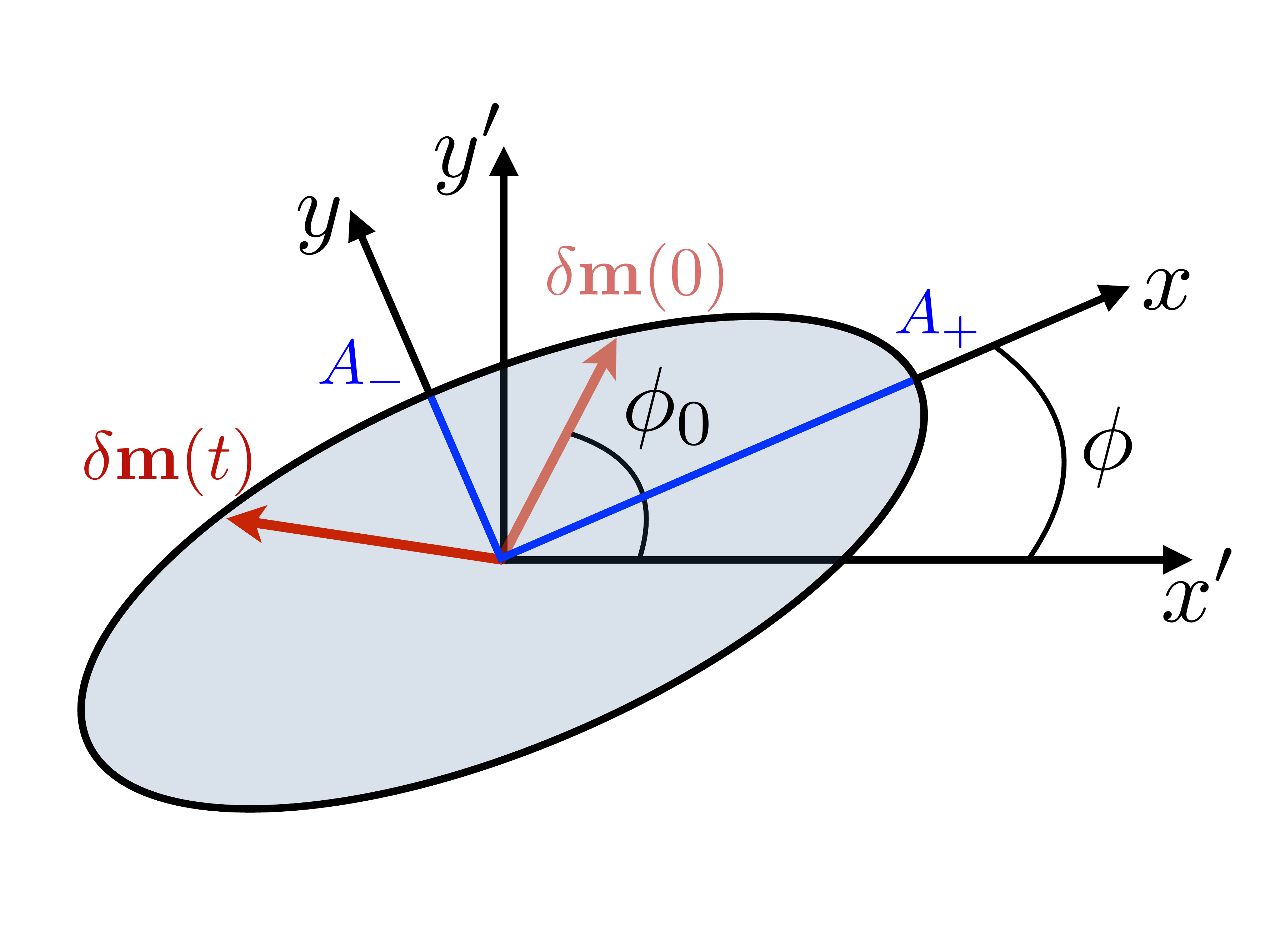}
 \caption{Schematic diagram of the elliptical precession. The time-dependent transverse magnetization vector $\delta \mathbf{M}$, illustrated in red, performs an elliptical motion starting at an azimuthal angle $\varphi=\phi_0$ for $t=0$. The ellipse has major and minor semiaxes given by $A_{\pm}=M\delta\theta(\cos\xi\pm\sin\xi)/2$ and is rotated by an angle $\phi$ with respect to the $x'$ axis.}
 \label{fig:ellipse}
\end{figure}

We first begin by writing the time-dependent magnetic moment in the local frame of reference as $\mathbf{M}(t) = M\cos\theta(t)\VEC{\hat{n}}_z+M\sin\theta(t)\left[\cos\varphi(t)\VEC{\hat{n}}_x+\sin\varphi(t)\VEC{\hat{n}}_y\right]$, where $\theta,\varphi$ are the usual spherical angles.
Assuming small deviations from the equilibrium orientation, this becomes $\mathbf{M}(t) \simeq M\VEC{\hat{n}}_z+\delta\mathbf{M}(t)$, with the small transverse components $\delta\mathbf{M}(t) = M\delta\theta(t)\left[\cos\varphi(t)\VEC{\hat{n}}_x+\sin\varphi(t)\VEC{\hat{n}}_y\right]$.
We then re-express $\delta\mathbf{M}(t)$ using the circular components of Eq.~\eqref{changebasis}, parametrized as $\delta M_- = M\delta\theta e^{\iu(\phi_0-\phi)}\cos\xi$ and $\delta M_+ = M\delta\theta e^{\iu(\phi_0+\phi)}\sin\xi$ (see Appendix~\ref{apx:ellipticalmode}).
Here, $\delta\theta$ is the small opening angle of the precessional cone, and $\xi$ sets the aspect ratio of the ellipse: Its semiaxes are $A_{\pm}=M\delta\theta(\cos\xi\pm\sin\xi)/2$.
Referring to Fig.~\ref{fig:ellipse}, the tilt of the ellipse away from the $x'$ axis is given by $\phi$ and $\phi_0$ is the initial phase of the motion.
The main parameters of the ellipse can then be obtained from the circular components as
%


\begin{equation}
\begin{split}
A_{\pm} =& \frac{|\delta M_-|\pm|\delta M_+|}{2} \\
\phi =& \frac{1}{2}\arg\left(\frac{\delta M_+}{\delta M_-}\right)
\end{split}\quad.
\end{equation}

It is instructive to consider the simple scenario where $\phi=0$, i.e., when the axes of the ellipse are aligned with the $x'$ and $y'$ directions. 
For $\omega>0$, the different kinds of precessional motion are: a counterclockwise circular polarization ($\circlearrowleft$) for $\xi=0$, an $x$ linear polarization ($\leftrightarrow$) for $\xi=\pi/4$, a clockwise circular polarization ($\circlearrowright$) for $\xi=\pi/2$, and a $y$ linear polarization ($\updownarrow$) for $\xi=3\pi/4$. 
Values of $\xi$ between these angles represent ellipses with different eccentricities.


Imagine that the magnetic system is disturbed by a counterclockwise circularly polarized magnetic field with angular frequency $\omega$.
Then, the components of the dynamical magnetization are given by Eq.~\eqref{suscmatrix}, and so the semiaxes of the ellipse and the tilt angle can then be obtained from the appropriate susceptibilities as
\begin{equation}\label{semiaxes}
\begin{split}
A_{\pm} =& \delta B_0^\text{ext}\left(\frac{|\chi_{-+}|\pm|\chi_{++}|}{2}\right)\\
\phi =& \frac{1}{2}\arg\left(\frac{\chi_{++}}{\chi_{-+}}\right)
\end{split}\quad.
\end{equation}
We shall characterize the elliptical motion by focusing on its frequency-dependent amplitude $A$ and eccentricity $e$ defined as
\begin{equation}\label{amplitude}
\begin{split}
A(\omega) =& \sqrt{\frac{A_+^2(\omega) + A_-^2(\omega)}{2}}\\
e(\omega) =& \sqrt{1-\frac{|A_-(\omega)|^2}{|A_+(\omega)|^2}}
\end{split}\quad.
\end{equation}
Circular precession is described by $e=0$, while the motion describes a linear oscillation for $e=1$.

%

In the following sections, we study the results obtained from first principles calculations, by making use of the analytical phenomenological expressions and the generic motion quantities detailed above, to describe the magnetization dynamics of single Fe adatoms and dimers deposited on a Cu(111) surface.

\section{Single magnetic impurity}
\label{sec:adatom}

\begin{figure}[!htb]
 \centering
  \includegraphics[width=0.5\columnwidth]{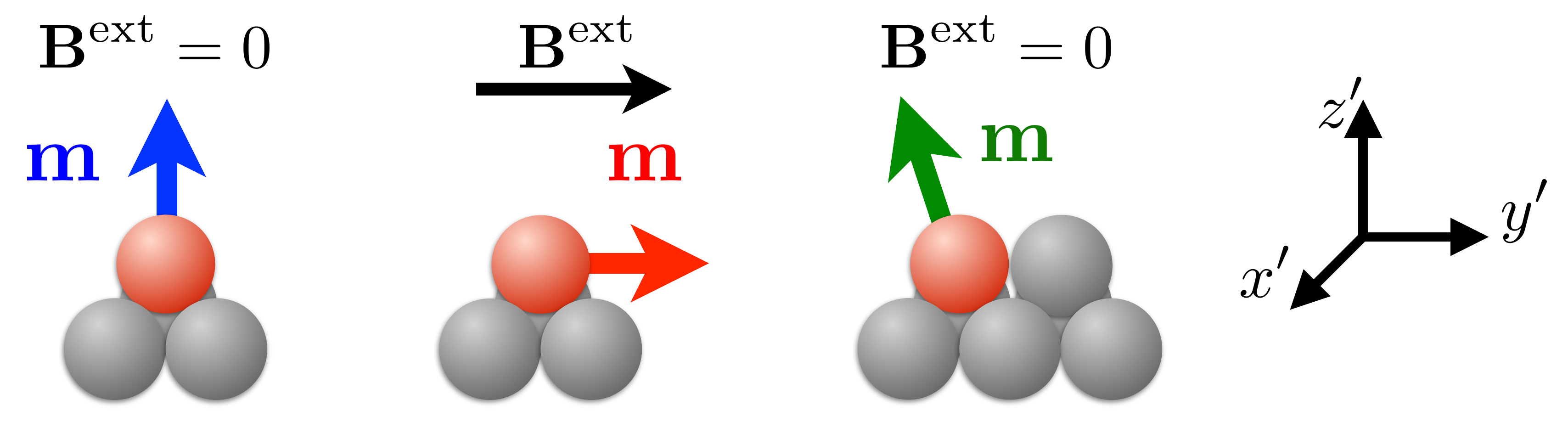}
 \caption{Diagrams of the three different Fe adatoms configurations deposited on Cu(111). Single Fe adatom without external field (blue), single Fe adatom with large field along $\VEC{\hat{n}}_y'$ (red), and Fe-Cu dimer (green). On the right, we display the global frame of reference.}
 \label{fig:adatoms}
\end{figure}

We start our investigation with a Fe adatom in three different structures, as depicted in Fig.~\ref{fig:adatoms}, and how to describe their dynamics.

\subsection{Out-of-plane magnetization}

\begin{figure}[!htb]
 \centering
  \includegraphics[width=0.5\columnwidth]{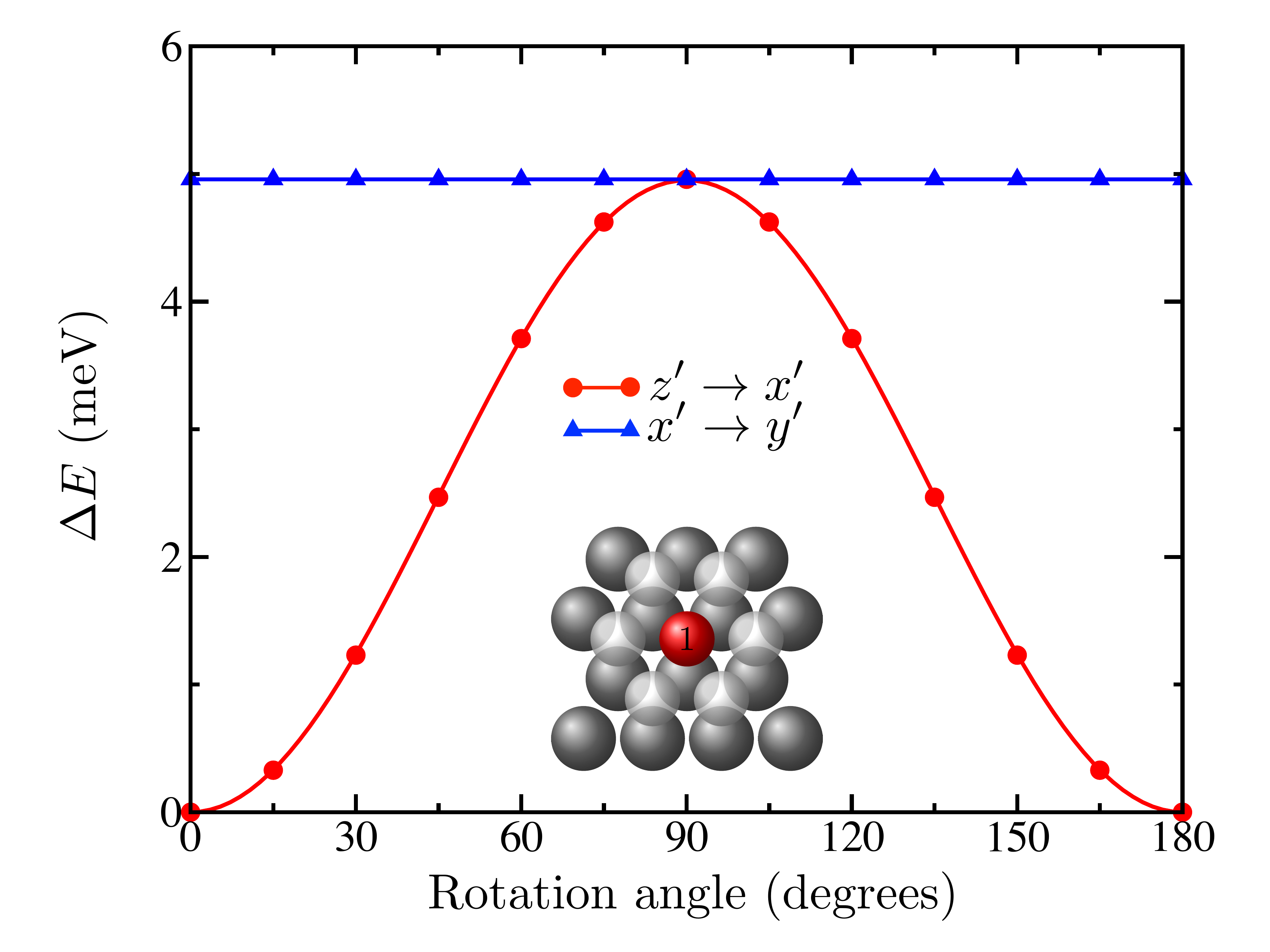}
 \caption{
Variation of the band energy for a single Fe atom on the Cu(111) surface.
The energy is plotted as a function of the magnetization direction, taking as a reference the self-consistent calculation where $\mathbf{M}\|\VEC{\hat{n}}_{z'}$, i.e., $\Delta E = E^\text{band}(\mathbf{M}) - E^\text{band}(M\VEC{\hat{n}}_{z'})$.
The system is schematically illustrated in the central diagram, with dark gray spheres representing Cu atoms, while transparent spheres are vacuum positions around the Fe adatom (red). 
For such $C_{3v}$ symmetry, the uniaxial anisotropy constant $K^{\Delta E}=\SI{4.95}{\milli\electronvolt}$ defined by the model Hamiltonian given in Eq.~\eqref{singleenergy} can be obtained directly from the variation of the energy when the magnetization is rotated from out-of-plane to in-plane, in the $z'x'$ plane (red circles).
There is no energy variation when the magnetization is rotated in the ($x'y'$) surface plane (blue triangles), confirming the uniaxial character of the anisotropy.}
 \label{fig:Fe_mae}
\end{figure}

First, consider a single magnetic Fe adatom deposited on the Cu(111) surface.
The energy of the system can be mapped into a simple model given by Eq.~\eqref{singleenergy}, where the spin magnetic moment is $M=3.2\mu_\text{B}$ while the orbital magnetic moment is $M^\text{orb}=0.55\mu_\text{B}$.
Due to the $C_{3v}$ symmetry of this system, it presents uniaxial anisotropy with $K_x = K_y = K$. 
In Fig.~\ref{fig:Fe_mae}, we show the band energy variation with respect to a self-consistent calculation for $\mathbf{M}\|\VEC{\hat{n}}_{z'}$, i.e., $\Delta E = E^\text{band}(\mathbf{M}) - E^\text{band}(M\VEC{\hat{n}}_{z'})$, as a function of the magnetization direction. 
Following the magnetic force theorem \cite{Weinert:1985ki}, we find a magnetic anisotropy energy constant of $K^{\Delta E} =\SI{4.95}{\milli\electronvolt}$, which corresponds to an easy axis uniaxial anisotropy, with the easy axis being the normal to the surface. Throughout the text, we use the superscript $\Delta E$ to indicate values obtained from band energy variations.

\begin{figure}[!htb]
 \centering
  \includegraphics[width=0.5\columnwidth]{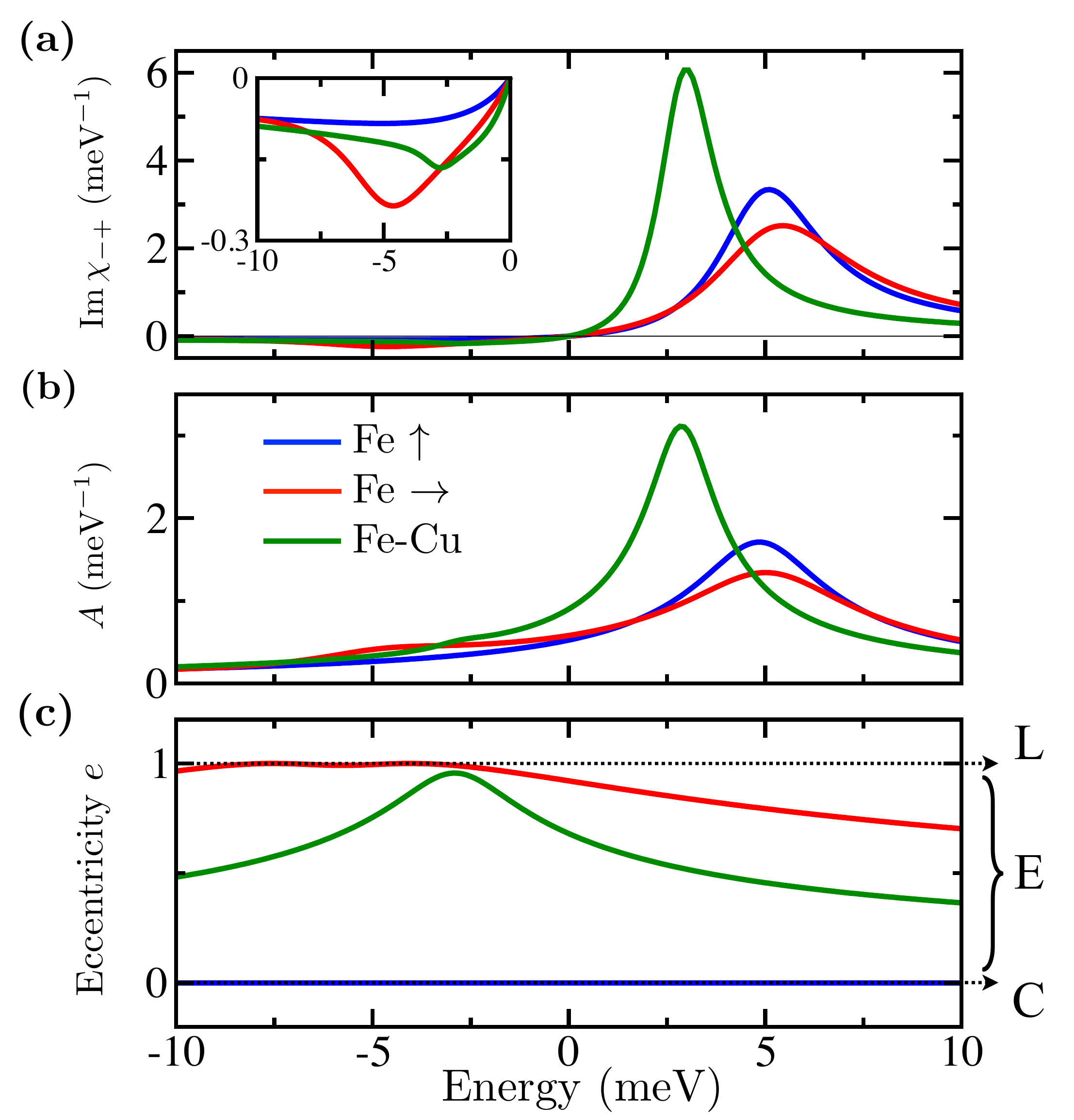}
 \caption{Dynamical excitations of single magnetic impurities. 
 (a) Imaginary part of the transverse dynamical susceptibility $\chi_{-+}$ as a function of the energy. The inset displays a magnification of the responses for negative energies.
 (b) and (c) illustrate the amplitude $A$ and eccentricity of the precession $e$, as defined in Eq.~\eqref{amplitude}, as a function of the energy when a circularly polarized magnetic field is applied.
L, E, and C indicate the regions with linear, elliptical, and circular oscillations.
 The curves for the three different magnetic adatom structures are color coded as follows: isolated Fe adatom (blue), isolated Fe adatom with in-plane magnetic field (red), Fe adatom with a neighboring Cu adatom (green). 
 See Fig.~\ref{fig:adatoms} for an illustration.
}
 \label{fig:chipm_adatom}
\end{figure}

We imagine that an STM tip is placed above the Fe adatom and that the tunneling electrons induce spin-flip excitations locally, which then lead to an inelastic contribution to the tunneling current \cite{Schweflinghaus:2014eq}.
We associate the dynamical spin excitations driven by this process with the local susceptibility $\chi_{-+}$.
For the $C_{3v}$ symmetry, the circular components of the effective field simplify to $\delta B_{\pm}^\text{eff} =-2K\delta M_{\pm}/M^2+\delta B_{\pm}^\text{ext}$.
The equations for $\delta M_{+}$ and $\delta M_{-}$ decouple and the susceptibility matrix given in Eq.~\eqref{suscmatrix} has only diagonal components
\begin{equation}\label{chi_adatom}
\begin{split}
\chi_{-+}(\omega) = \frac{\delta M_-}{\delta B^\text{ext}_-} =& \frac{-\gamma_+M}{\left(\omega-\gamma_+B_\|\right)}\quad,
\end{split}
\end{equation}
and $\chi_{+-}(\omega) = \left[\chi_{-+}(-\omega)\right]^*$, where $\gamma_+ = \frac{\gamma}{1+ \iu\alpha}$ and $B_\| = B^\text{ext}_z +2K/M$.
Note that each susceptibility has a single pole, located at $\omega^{0} = \gamma_+B_\|$ and $\omega^{0} = -\gamma^*_+B_\|$, respectively.
Figure~\ref{fig:chipm_adatom}(a) (blue curve) shows the density of spin excitations from the TDDFT calculation (obtained from $\operatorname{Im}\chi_{-+}$) when no static magnetic field is applied to the sample.
This quantity is related to the steplike features in the tunneling conductance measured in ISTS experiments \cite{Schweflinghaus:2014eq}.

The values of the anisotropy constant $K$, the effective gyromagnetic ratio $\gamma$, and the damping parameter $\alpha$ can be extracted by fitting the linear behavior of the real and imaginary part of $[\chi_{-+}]^{-1}$, shown in Eq.~\eqref{adatom_uniaxial+-}, close to $\omega=0$ \cite{dosSantosDias:2015bh}.
These values are listed in Table~\ref{tab:parameters}.
The fitted values of $K$ differ slightly from the ones calculated using the band energy variation --- while the latter captures the energy difference between two orthogonal directions of the magnetic moment, the former represents the curvature of the energy around its equilibrium direction.
Note, however, that the gyromagnetic factor $\gamma$ shifts considerably from the expected value of $2$.
This induced delay occurs due to the high hybridization with the surface \cite{dosSantosDias:2015bh}, which leads to large spin currents pumped out of the adatom \cite{Tserkovnyak:2002ju}.
This strong hybridization is also responsible for the high value obtained for the Gilbert damping $\alpha$.
The shift of the resonance energy to imaginary values, $\operatorname{Im}\gamma_+B_\|=-\frac{\alpha\gamma B_\|}{1+\alpha^2}$, is proportional to the damping parameter and is responsible for the broadening of the response functions seen in Fig.~\ref{fig:chipm_adatom}(a).
This effect can also be seen from the time dependence of the transverse magnetization components obtained in Eq.~\eqref{transvmag}, where this contribution appears as an exponential decay of their amplitude.


\begin{center}
\begin{table}[ht!]
\caption{Parameters from the ground state DFT calculations (upper half) and from fitting the TDDFT susceptibility (lower half) for the three different Fe adatom structures deposited on Cu(111) investigated: single Fe adatom without external field, single Fe adatom with large field along $\VEC{\hat{n}}_{y'}$, Fe-Cu dimer.
For the adatom in the presence of magnetic field, $\alpha$ is highly anisotropic: $\alpha_{xx}=0.34$, while $\alpha_{yy}=0.39$, in the local frame of reference.}
\begin{center}
\begin{tabular}{c|c|c|c}
\hline\hline
 & \thead{Fe adatom \\(no $\mathbf{B}^\text{ext}$)\\$\uparrow$} & \thead{Fe adatom \\ ($\mathbf{B}^\text{ext})$\\$\rightarrow$} & \thead{Fe-Cu dimer \\(no $\mathbf{B}^\text{ext}$)\\$\nwarrow$} \\\hline
$M$ ($\mu_\text{B}$)                                 & 3.24  & \phantom{-}3.24\phantom{*} & 3.19 \\
$M^\text{orb}$ ($\mu_\text{B}$)                 &  0.55 & \phantom{-}0.24\phantom{*} & 0.41 \\
$K^{\Delta E}_x$ (\SI{}{\milli\electronvolt}) & 4.95 & \phantom{-}0.02\phantom{*} & 2.48 \\
$K^{\Delta E}_y$ (\SI{}{\milli\electronvolt}) & 4.95 & -5.46\phantom{*}                  &  3.40 \\\hline
$K_x$ (\SI{}{\milli\electronvolt})                  & 4.98 & \phantom{-}0.07\phantom{*} & 2.49 \\
$K_y$ (\SI{}{\milli\electronvolt})                  & 4.98 & -5.43\phantom{*}                  & 3.37 \\
$\gamma$                                                  &1.72  & \phantom{-}1.71\phantom{*} & 1.71 \\
$\alpha$                                                     & 0.33  & \phantom{-}0.39*                  & 0.29 \\
\hline\hline
\end{tabular}
\end{center}
\label{tab:parameters}
\end{table}%
\end{center}

Since the susceptibility matrix is diagonal, the amplitude of oscillation $A(\omega) = |A_+(\omega)| = |A_-(\omega)|$, and the eccentricity $e=0$ for every frequency.
These quantities are shown in the blue curves of Figs.~\ref{fig:chipm_adatom}(b) and (c), and describe a circular precessional motion --- reflecting an effective field with equal transverse components acting on the magnetization.
The maximum amplitude of oscillation presented in Fig.~\ref{fig:chipm_adatom}(b) is close to the maximum density of spin excitations obtained in Fig.~\ref{fig:chipm_adatom}(c), although the former presents a much broader peak due to contribution of the real part of $\chi_{-+}$ in this quantity, see Eq.~\eqref{semiaxes}.


\subsection{In-plane magnetization}

In the presence of spin-orbit coupling, the symmetry of the system can be broken with external magnetic fields. 
By applying a large $\VEC{B}^\text{ext}$ along the $\VEC{\hat{n}}_{y'}$ direction of the global frame of reference, we rotate the magnetization of the adatom to lie in the surface plane along this direction (which defines the local $\VEC{\hat{n}}_z$). 
We chose a static magnetic field of \SI{90}{\tesla} such that the spin excitation resonance is similar to the one obtained for the out-of-plane magnetization.
Although this represents an artificially large magnetic field, it enables an easy comparison between both cases.

The strong magnetic field has only a minor impact on the ground state properties of the adatom.
The spin moment is unchanged, but the orbital moment is less than half of the value of the out-of-plane case.
This reduction occurs as the spin moment is now perpendicular to the $C_{3v}$ symmetry axis, which, due to the spin-orbit coupling, leads to a lowering of symmetry.
In the local frame of reference, the magnetic anisotropy matrix can now be described by Eq.~\eqref{kmatrix} with $K_x=0$, $K_y=-K$.

The density of spin excitations is depicted by the red curve of Fig.~\ref{fig:chipm_adatom}(a).
For positive frequencies, the curves are very similar with a slightly larger broadening when the magnetization points in the plane.
A more detailed picture is provided by the parameters listed in Table~\ref{tab:parameters}, obtained from the fits of the different components of the inverse susceptibilities to Eq.~\eqref{adatom_uniaxial_bext+-}.
The shift of the gyromagnetic factor away from $2$ is very similar to the one obtained before, indicating that the spin pumping mechanism is essentially isotropic.
In contrast, the damping parameter is substantially anisotropic, with $(\alpha_{xx}-\alpha_{yy})/\alpha_{yy} \sim 13\%$, where $\alpha_{\mu\mu}$ was obtained from the $\chi_{\mu\mu}^{-1}$ component of the susceptibility.

Surprisingly, there appears to be a new excitation peak manifested at negative energies.
Even though the value of the response function is small, it represents a fundamental difference in the spin dynamics and in the precessional shape.
It originates from the anisotropic effective field along the transversal directions.
This can be understood from the expression for the transverse susceptibility matrix, Eq.~\eqref{adatom_uniaxial_bext+-}.
We find
\begin{equation}\label{chi+-bext}
\begin{split}
\chi_{-+} 
=&\frac{\gamma_+M\,\big(\omega+\gamma^*_+B_\|^*\big)}{\omega^+ - \omega^-}\left(\frac{1}{\omega-\omega^-}-\frac{1}{\omega-\omega^+} \right)
\end{split}\quad.
\end{equation}
where the components of the effective field are $B_\| = B^\text{ext} - K/M$ and $B_\perp = K/M$, defining the poles
\begin{equation}\label{eigenenergies}
\begin{split}
\omega^\pm = \iu\operatorname{Im}(\gamma_+B_\|)\pm\sqrt{ [\operatorname{Re}(\gamma_+B_\|)]^2-|\gamma_+B_\perp|^2 }
\end{split}\quad.
\end{equation}
This result explains why, besides the resonance at $\omega^+$, there is also a signal at $\omega^-$, as seen in the red curve of the inset of Fig.~\ref{fig:chipm_adatom}(a). 
For the uniaxial case ($B_\perp = 0$), the second peak is absent because the numerator is proportional to $\omega+\gamma_+^*B_\|^* = \omega-\omega^-$ and so the first term in Eq.~\eqref{chi+-bext} becomes a constant.
Furthermore, the precession becomes circular.
We can then distinguish circular and elliptical precession directly from $\chi_{-+}$, by inspecting how many resonances it displays.
In addition, $B_\perp$ reflects the ellipticity of the transverse effective field, and it lowers the resonance frequency in comparison to the uniaxial case.
As before, the imaginary part obtained in Eq.~\eqref{eigenenergies} is proportional to the damping parameter $\alpha$ and is responsible for the decay of the magnetization back into the equilibrium direction.

As described in Appendix~\ref{apx:ellipticalmode}, even though there are two resonant energies of the system, they both represent a single elliptical precession.
The amplitude of the motion followed by the transverse component of the magnetization is illustrated in the red curve of Fig.~\ref{fig:chipm_adatom}(b).
Although the amplitude is smaller than the out-of-plane case at the resonance, the peak is broader and the response is larger for negative frequencies close to $\omega=\omega^-$.
In this region, the oscillation is close to linear, as evidenced by the eccentricity close to $e=1$ displayed in Fig.~\ref{fig:chipm_adatom}(c).
For higher frequencies, the precession amplitude and eccentricity decays, approaching a circular shape.


For this last scenario, we applied a large magnetic field to rotate the magnetization and make use of different anisotropy energies along the transverse directions to induce elliptical motion of the magnetization precession.
In the next section, we show how this can be achieved by manipulating the surroundings of the magnetic unit.

\subsection{Engineering magnetic anisotropy}
\label{sec:FeCudimer}

\begin{figure}[!htb]
 \centering
  \includegraphics[width=0.5\columnwidth]{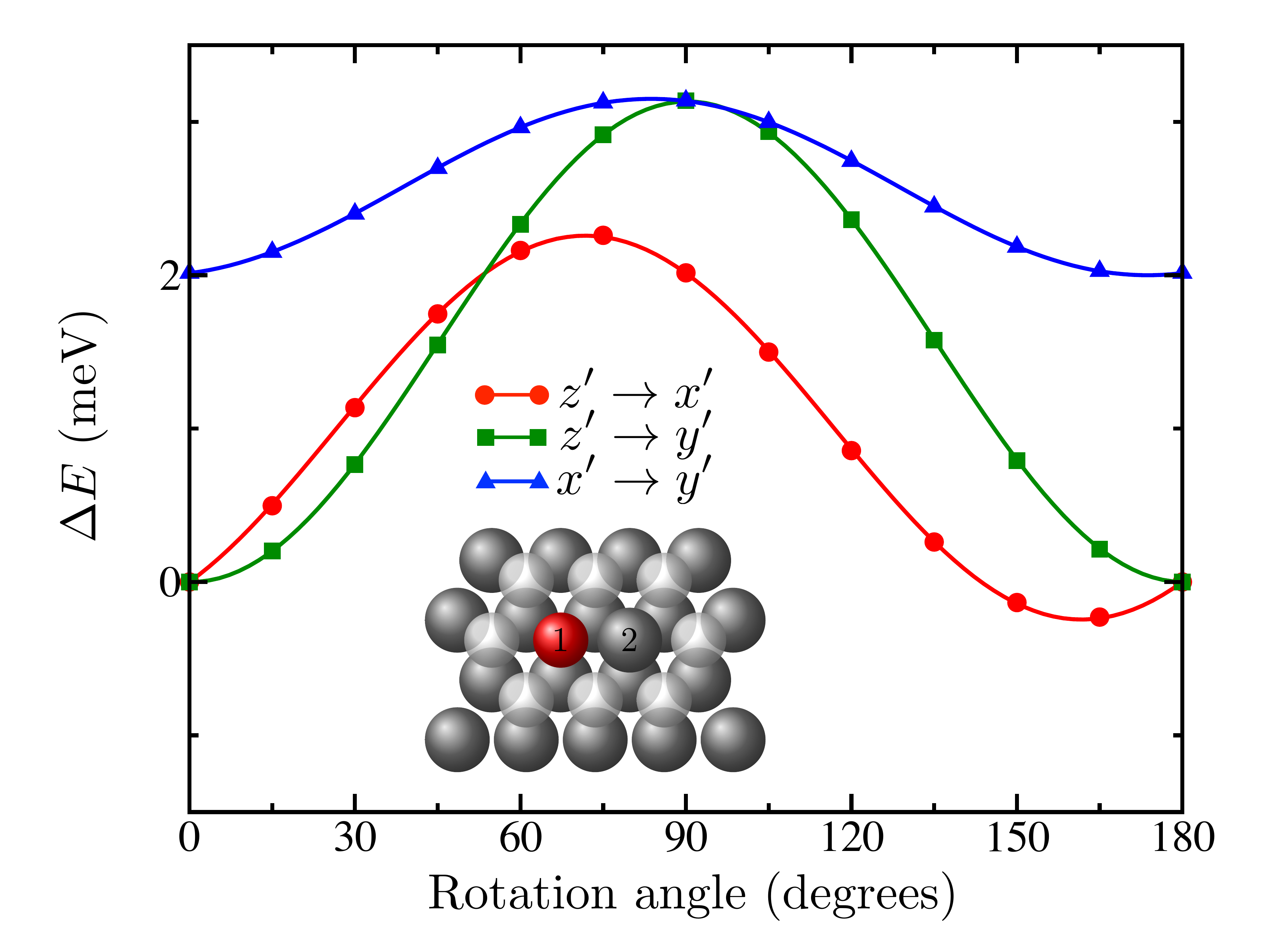}
 \caption{
 Variation of the band energy for a Fe-Cu dimer on the Cu(111) surface.
 The energy is plotted as a function of the magnetization direction, taking as a reference the self-consistent calculation where $\mathbf{M}\|\VEC{\hat{n}}_{z'}$, i.e., $\Delta E = E^\text{band}(\mathbf{M}) - E^\text{band}(M\VEC{\hat{n}}_{z'})$.
 The system is schematically illustrated in the central diagram, with dark gray spheres representing Cu atoms and transparent spheres are vacuum positions around the Fe (1) and Cu (2) dimer. 
The anisotropy constants defined for the model Hamiltonian given in Eq.~\eqref{singleenergy} are obtained by fitting the energy curves when the magnetization is rotated along the $z'x'$ plane (red circles), the $z'y'$ plane (green squares), and along the $x'y'$ plane (blue triangles). 
The values of the magnetic anisotropy components in the local frame of reference of the ground state magnetization are listed in Table~\ref{tab:parameters}.}
 \label{fig:FeCu_mae}
\end{figure}

As an alternative to an external magnetic field, we now explore the impact of a neighboring nonmagnetic Cu atom on the spin excitations of the Fe atom.
The rotational symmetry is broken in real space and, due to the spin-orbit coupling, also in spin space, resulting in a tilt of the equilibrium direction of the spin moment away from the surface normal.
The spin and orbital magnetic moments decrease compared to the out-of-plane case, reaching $M=3.19\MUB$ and $M^\text{orb}=0.41\MUB$.
The latter represent a 25\% decrease in the orbital moment, which indicates a strong influence from the additional atom.
The symmetry lowering is confirmed by the variation of band energy of the system when the magnetization angle is rotated along the three different axes, illustrated in Fig.~\ref{fig:FeCu_mae}.
The intricate energy landscape leads to a tilt of the spin moment of $\theta \sim 18^\circ$ and $\phi\sim 177^\circ$, in spherical coordinates, close to the $x'z'$ plane.
In the local frame of reference, $K_x^{\Delta E} = \SI{2.48}{\milli\electronvolt}$ and $K_y^{\Delta E} = \SI{3.40}{\milli\electronvolt}$, which corresponds to a biaxial anisotropy, with the magnitude of the anisotropy constants reduced from the isolated Fe adatom case.
Extra Cu atoms decrease their values even more.

This is also reflected on the spin excitation spectra, green curve in Fig.~\ref{fig:chipm_adatom}(a), in which the resonance is shifted to lower frequencies, leading also to a higher amplitude of precession, Fig.~\ref{fig:chipm_adatom}(b).
The phenomenological parameters are again obtained by fitting the first-principles results to Eq.~$\eqref{atadom_biaxial+-}$, with the values
listed in Table~\ref{tab:parameters}.
The gyromagnetic factor and the damping parameter are very similar to the out-of-plane adatom case, but the biaxial anisotropy is manifested in the elliptical precession indicated by the peak at negative energies [inset of Fig.~\ref{fig:chipm_adatom}(a)].
It is also seen from the eccentricity plotted in Fig.~\ref{fig:chipm_adatom}(c) that reaches close to $e=1$ --- linear oscillation --- at $\omega=\omega^-$ and decreases away from this frequency.
Our results demonstrate that atomic manipulation not only of the magnetic unit but also of its environment can be used to control and tune possible magnetic excitations and may be used to operate static and dynamical aspects of nanostructures.

\section{Magnetic dimers}
\label{sec:dimer}

\begin{figure}[!htb]
 \centering
  \includegraphics[width=0.5\columnwidth]{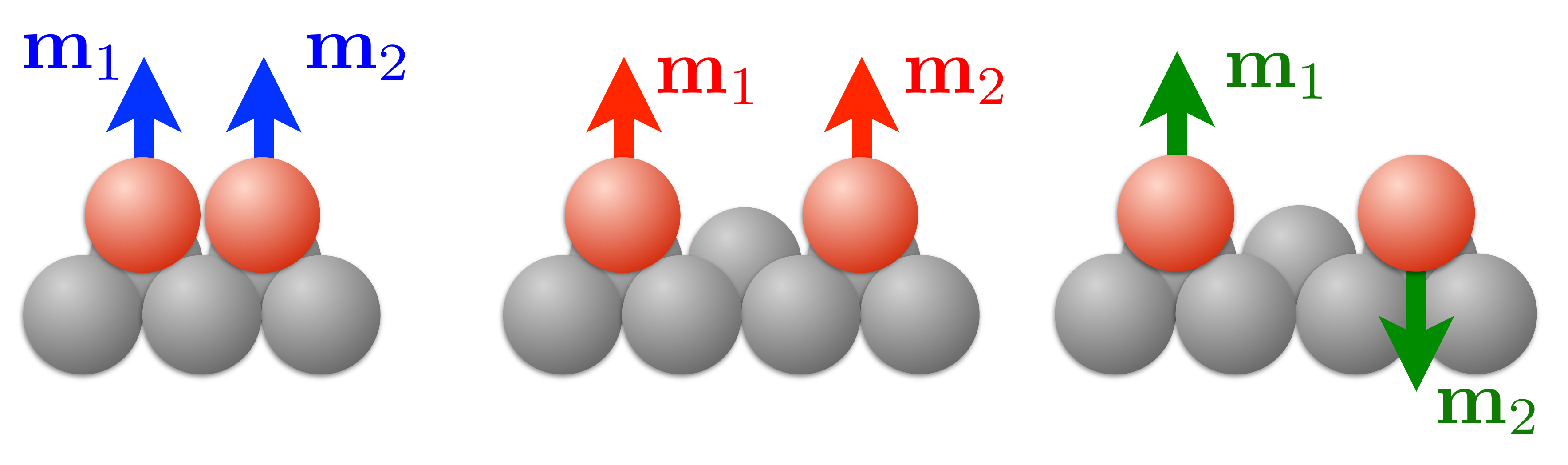}
 \caption{
Diagrams of the three different Fe dimers deposited on Cu(111). Nearest neighbor distance with large ferromagnetic coupling (blue); twice the nearest neighbor distance with small antiferromagnetic coupling, starting from a ferromagnetic state (red) or from an antiferromagnetic state (green).
 }
 \label{fig:dimers}
\end{figure}

We focus now on magnetic dimer structures and how their ground state properties affect the possible excitation modes of the system. 
The energy of these structures may be mapped into a model given by Eq.~\eqref{energy}.
By manipulating the separation between the Fe adatoms and taking advantage of the distance-dependent oscillatory exchange interaction between the adatoms, we design two different structures:
\begin{itemize}
\item Dimer I --- adatoms separated by the nearest-neighbor distance on the surface have a strong ferromagnetic coupling.
\item Dimer II --- adatoms separated by twice the nearest-neighbor distance on the surface have a weak antiferromagnetic coupling.
\end{itemize}
For dimer II, we investigate excitations starting from the metastable FM state and from the AFM ground state.
The three cases we consider are depicted schematically in Fig.~\ref{fig:dimers}.

When bringing two adatoms together to form a dimer, the symmetry is lowered from $C_{3v}$ to $C_s$, leaving only one mirror plane.
The system presents a biaxial anisotropy that can be mapped into the $\mathbf{K}$ matrix given by Eq.~\eqref{kmatrix}.
The values of $K_x^{\Delta E}$ and $K_y^{\Delta E}$ (per Fe atom) are obtained from the change of the band energy when the magnetic moments are rotated simultaneously along the different directions.

In dynamical studies of nanostructured systems, a central role is played by the coupling between the magnetic moments.
For the dimers considered in this paper, the spin-orbit interaction does not lead to appreciable anisotropic pair interactions, such as the Dzyaloshinskii-Moriya interaction.
Their values, obtained from the calculated susceptibilities, are two orders of magnitude smaller than the exchange interaction --- as one would expect for systems with low spin-orbit coupling such as Cu.
Hence, we consider only the simpler isotropic Heisenberg exchange in our phenomenological model.
As there are several ways of estimating $J$, we first provide an overview of the methods before analyzing each dimer structure.

\subsection{Exchange coupling}

A first definition of $J$ is given by assuming that the Heisenberg coupling appropriately describes the energetics of our dimers. 
Identifying the first-principles total energy difference between the antiferro- and the ferromagnetic states with the value expected from the model we find
\begin{equation} \label{eq:Dimer_J_from_DeltaE}
2J^{\Delta E} = E^\text{AFM}-E^\text{FM}
\quad.
\end{equation}
In this convention, a positive (negative) value of $J^{\Delta E}$ favors a ferromagnetic (antiferromagnetic) ground state. 

The mapping between the Heisenberg coupling and the total energy difference of the two magnetic states assumes that the coupling constant (and the electronic structure) is not substantially affected by the angle between the two magnetic moments --- which, in practice, may not be a reasonable assumption.
To avoid this problem, the exchange coupling may also be derived via infinitesimal rotation of the magnetic moments taking into account the magnetic force theorem~\cite{Liechtenstein:1984fj,Antropov:2003ii}, which can be written for ferro- and antiferromagnetic alignments as
\begin{equation}
  \label{eq:WCCu111_dimer_J0_chi_connection}
    J_0
  = \mp(M K_{\MR{T}}^{\MR{xc}})^2 \chi^{\text{KS}}_{1-,2+} \quad.
\end{equation}
See Sec.~\ref{sec:tddft} for a description of these quantities.

This result, however, does not take into account many-body effects. 
Here we introduce yet another method to obtain the coupling parameter, based on the mapping of the susceptibilities calculated in TDDFT to those obtained from the LLG model, Eqs.~\eqref{dimer_ferro+-} and \eqref{dimer_anti+-}. 
For ferromagnetic or antiferromagnetic ground states, one finds
\begin{equation}
  \label{eq:WCCu111_J_chi_connection}
\begin{split}
J &= \mp M^2\left[ \chi^{-1}\right]_{1-,2+}\quad,
\end{split}
\end{equation}
where $1$ and $2$ label the two magnetic moments for which the coupling constant is determined, and the sign $-$ ($+$) is used for the ferro- (antiferro-) magnetic alignment.

The expressions for the exchange couplings obtained in Eqs.~\eqref{eq:WCCu111_dimer_J0_chi_connection} and \eqref{eq:WCCu111_J_chi_connection} are related by
\begin{equation}
  \label{eq:WCCu111_dimer_J0_J_connection}
\begin{split}
  J &= J_0 \left[ 1 \pm \frac{2J_0}{M^2K_{\MR{T}}^{\MR{xc}}} \right]^{-1} \;,
\end{split}
\end{equation}
where $+$ and $-$ signs account for a FM and an AFM alignment of the two involved magnetic moments, respectively.
The form of Eq.~\eqref{eq:WCCu111_dimer_J0_J_connection} has been discussed in the literature~\cite{Bruno:2003ir,Katsnelson:2004ha}: Similar to the connection between $\chi$ and $\chi^\text{KS}$ via $K_{\MR{T}}^{\MR{xc}}$, the exchange coupling constant $J$ can be seen as a renormalization of $J_0$ by $K_{\MR{T}}^{\MR{xc}}$, the exchange-correlation kernel. 
Since the spin splitting is much larger than the spin excitation energies for the systems we investigate, we expect the values of $J_0$ and $J$ to be quite similar to each other \cite{Katsnelson:2004ha}.
In the following sections, we analyze the ground state and dynamical properties of the different dimer configurations and how the coupling constant and the initial magnetic configuration affect the excitation energies. 

\subsection{Dimer I}

\begin{figure}[!htb]
  \centering
   \includegraphics[width=0.5\columnwidth]{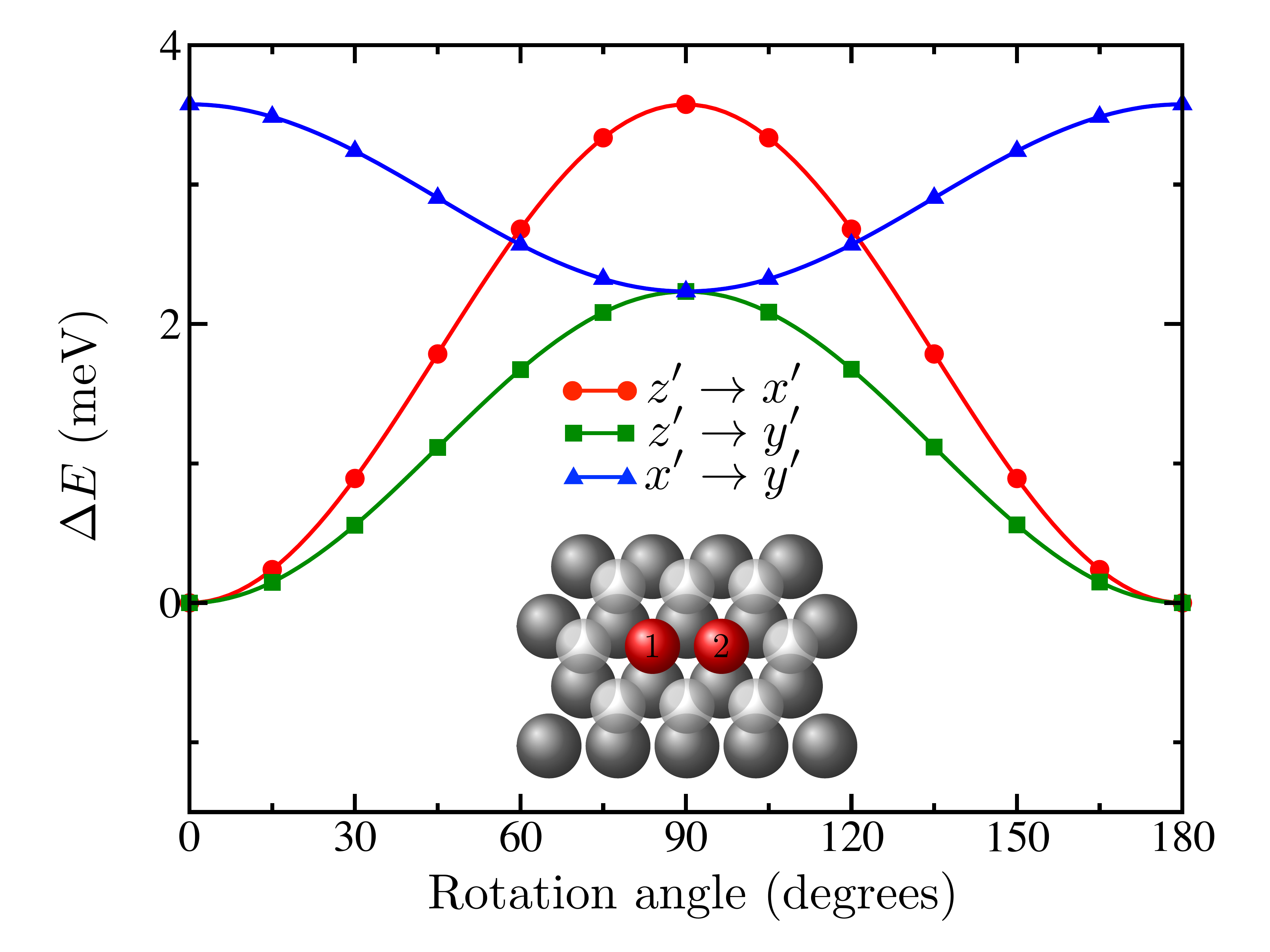}
 \caption{
  Variation of the band energy for the dimer I structure.
The energy is plotted as a function of the magnetization direction of the parallel alignment, $\mathbf{M} = (M_1+M_2)\VEC{\hat{n}}$, taking as a reference the self-consistent calculation where $\mathbf{M}\|\VEC{\hat{n}}_{z'}$, i.e., $\Delta E = E^\text{band}(\mathbf{M}) - E^\text{band}(M\VEC{\hat{n}}_{z'})$.
The dimer is composed by two nearest-neighbor Fe atoms (1 and 2) on the Cu(111) surface, as illustrated in the central diagram. 
Dark gray spheres represent Cu atoms and transparent spheres are vacuum positions around the nearest neighbor Fe-Fe dimer. 
The anisotropy constants are obtained by fitting the energy curves when $\mathbf{M}$ is rotated along the $z'x'$ plane (red circles), the $z'y'$ plane (green squares), and along the $x'y'$ plane (blue triangles).
The values of the magnetic anisotropy components in the local frame of reference are listed in Table~\ref{tab:parameters_dimer}.}
 \label{fig:FeFe_mae_nn}
\end{figure}

The spin and orbital magnetic moments per site, $M$ and $M^\text{orb}$ are given in Table~\ref{tab:parameters_dimer}.
Comparing with the isolated Fe adatom case (see Table~\ref{tab:parameters}), the spin moment decreases slightly while the orbital moment is drastically reduced by 65\%.
Due to the broken rotation symmetry in the plane, the system presents biaxial anisotropy --- as demonstrated by the band energy variation when the moments are rotated simultaneously along all directions, shown in Fig.~\ref{fig:FeFe_mae_nn}.
The anisotropy constants per atom are listed in Table~\ref{tab:parameters_dimer}.
Their magnitudes are 3-4 times smaller than for the isolated adatom and about half of those for the Fe-Cu dimer, revealing the impact of the strong hybridization between the $d$ orbitals of the two Fe atoms.

The strong $d$-$d$ hybridization is also evident in the large values of the ferromagnetic coupling $J$.
From the total energy difference, Eq.~\eqref{eq:Dimer_J_from_DeltaE}, we obtain $J^{\Delta E}=\SI{239}{\milli\electronvolt}$, while from the magnetic force theorem, Eq.~\eqref{eq:WCCu111_dimer_J0_chi_connection}, $J_0=\SI{193}{\milli\electronvolt}$.
This sizable difference reflects the change in the electronic structure going from the FM to the AFM state.

\begin{table}[ht!]
\centering
\caption{Parameters from the ground state DFT calculations (upper half) and from fitting the TDDFT susceptibility (bottom half) for the three Fe dimer structures deposited on Cu(111) investigated: nearest neighbor distance and twice the nearest neighbor distance with both ferromagnetic and antiferromagnetic alignments.
Values are given per atom.}
\begin{tabular}{c|c|c|c}
\hline\hline
& \thead{Dimer I\\$\uparrow\uparrow$}    & \thead{Dimer II \\$\uparrow\quad\uparrow$}   & \thead{Dimer II \\$\uparrow\quad\downarrow$} \\ \hline
$M$ ($\mu_\text{B}$)                                & \phantom{00}3.12  & \phantom{-}3.24 & \phantom{-}3.24 \\
$M^{\text{orb}}$ ($\mu_\text{B}$)             & \phantom{00}0.19  & \phantom{-}0.53 & \phantom{-}0.55 \\
$K^{\Delta E}_x$(\SI{}{\milli\electronvolt}) & \phantom{00}1.79 & \phantom{-}4.36 & \phantom{-}4.89 \\
$K^{\Delta E}_y$(\SI{}{\milli\electronvolt}) & \phantom{00}1.12 & \phantom{-}4.38  & \phantom{-}4.80 \\ 
$J^{\Delta E}$(\SI{}{\milli\electronvolt})     & 239\phantom{.00} & -3.9\phantom{0} & -3.9\phantom{0} \\
$J_0$ (\SI{}{\milli\electronvolt})                 & 193\phantom{.00} & -2.4\phantom{0} & -3.9\phantom{0} \\ \hline
$K_x$ (\SI{}{\milli\electronvolt})                 & \phantom{00}1.82 & \phantom{-}4.45  & \phantom{-}4.91 \\
$K_y$ (\SI{}{\milli\electronvolt})                 & \phantom{00}1.14 & \phantom{-}4.44  & \phantom{-}4.82 \\
$J$ (\SI{}{\milli\electronvolt})                     & 206\phantom{.00} & -2.8\phantom{0}  & -4.3\phantom{0} \\
$\gamma$                                                 & \phantom{00}1.85 & \phantom{-}1.69  & \phantom{-}1.71 \\
$\alpha$                                                    & \phantom{00}0.12 & \phantom{-}0.32  & \phantom{-}0.30
\\ \hline \hline
\end{tabular}
\label{tab:parameters_dimer}
\end{table}%

We also obtain the phenomenological parameters by fitting the results to the inverse susceptibility obtained in Eq.~\eqref{dimer_ferro+-}.
Their values are listed in Table~\ref{tab:parameters_dimer}.
The anisotropies are in very good agreement with the ones obtained by the band energy variation, and $J$ is very close to $J_0$, as expected.
We obtain relatively small values for the damping parameter, $\alpha=0.12$, which will be discussed below.
The gyromagnetic ratio $\gamma$ is closer to 2 than before, which indicates that the spin pumping mechanism is less efficient.

For a ferromagnetic dimer, we expect two precessional modes: a uniform mode where the magnetic moments precess in-phase (acoustic mode), and one mode where they precess with a phase difference of $\pi$ (optical mode).
By linearizing the equation of motion given in Eq.~\eqref{dimer_ferro+-}, the poles corresponding to the acoustic and optical mode are
\begin{equation}\label{eigenenergies_dimer_fm_alpha}
\begin{split}
\frac{\omega^{\pm}_\text{ac}}{\gamma'}=& -\iu\alpha(K_x+K_y)\pm\sqrt{4K_xK_y-\alpha^2(K_x-K_y)^2}\\
\frac{\omega^{\pm}_\text{op}}{\gamma'}=& -\iu\alpha(K_x+K_y+2J)\\
&\pm\sqrt{4(K_x+J)(K_y+J)-\alpha^2(K_x-K_y)^2}
\end{split}\quad,
\end{equation}
where $\gamma' = \frac{\gamma}{M(1+\alpha^2)}$. 
Since $J\gg K_x,K_y$, the optical mode is located at $2\gamma J/M \sim\SI{250}{\milli\electronvolt}$.
If $\omega \ll J$, the two spin moments stay parallel to each other, and the system behaves as a macrospin with biaxial magnetic anisotropy.
In this case, $\chi_{-+}$ can be described by Eq.~\eqref{chi+-bext} with $B_\| = (K_x+K_y)/M$ and $B_\perp = (K_x-K_y)/M$, and the excitations follow similar dynamics as the adatom with magnetization in-plane and the Fe-Cu dimer described in Sec.~\ref{sec:adatom}.
The expressions for the acoustic and optical frequencies, Eq.~\eqref{eigenenergies_dimer_fm_alpha}, show that in biaxial systems the damping may play an important role, lowering the resonance frequency.
These results are not captured by the so-called Kittel's formula \cite{Farle:1998gz}, which neglects damping effects.

\begin{figure}[!htb]
 \centering
  \includegraphics[width=0.5\columnwidth]{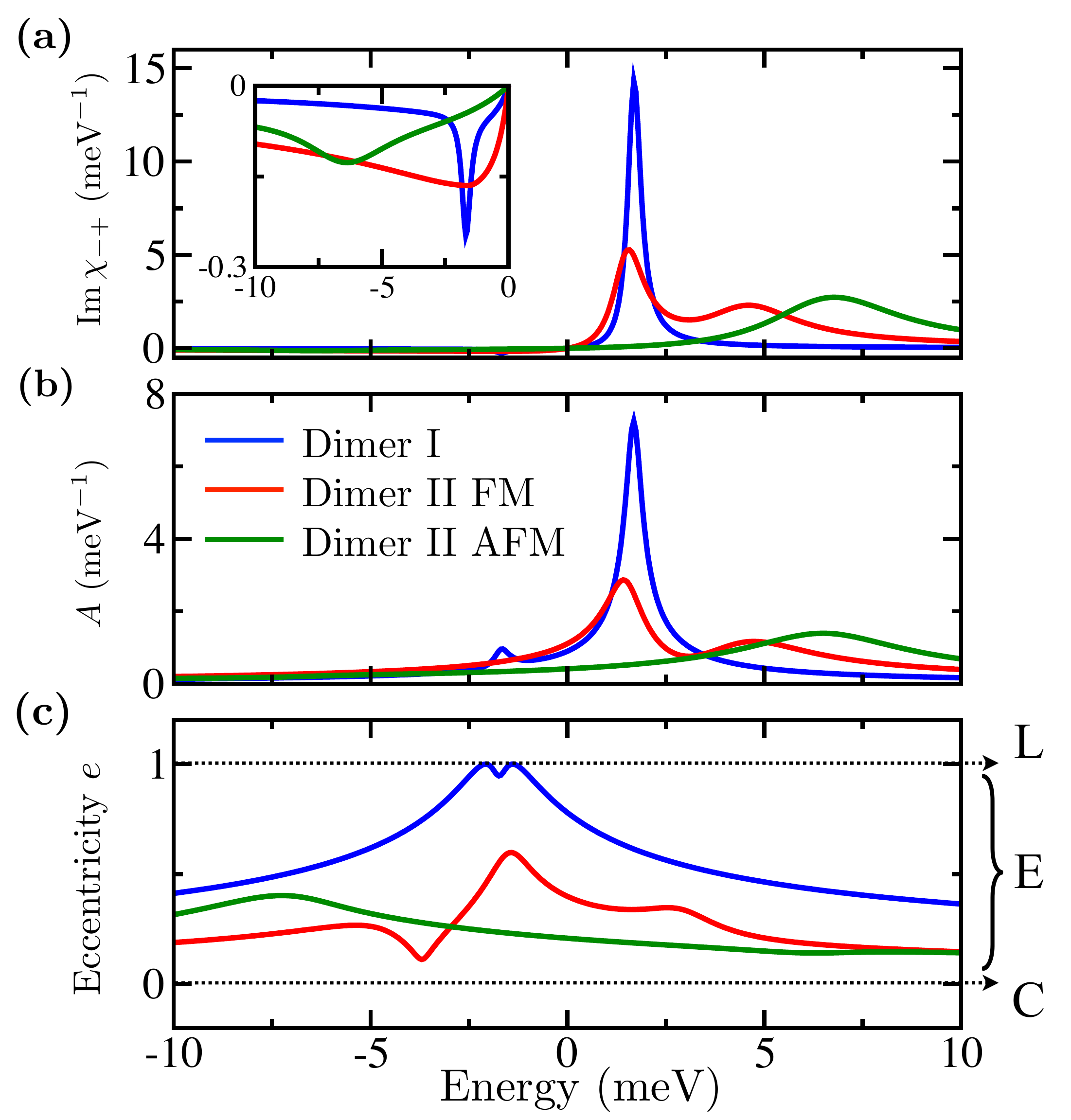}
 \caption{Dynamical excitations of dimers. 
 (a) Imaginary part of the local transverse dynamical susceptibility $\chi_{-+}$ as a function of the energy. 
 The inset displays a magnification of the responses for negative energies.
 (b) and (c) illustrates the amplitude $A$ and eccentricity of the precession $e$, as defined in Eq.~\eqref{amplitude}, as a function of the energy when a circularly polarized magnetic field is applied.
 L, E, and C indicate the regions with linear, elliptical, and circular oscillations.
 The curves for the three different dimer structures are color coded as follows: dimer I (blue), dimer II with FM alignment (red), dimer II with AFM alignment (green). See Fig.~\ref{fig:dimers} for an illustration.
 }
 \label{fig:chipm_dimers}
\end{figure}

We imagine that an STM tip is placed above atom 1 and that the tunneling electrons induce spin-flip excitations locally, which then lead to an inelastic contribution to the tunneling current \cite{Schweflinghaus:2014eq}.
We associate the dynamical spin excitations driven by this process in atom 1 with the local susceptibility $\chi_{1-,1+}\equiv\chi_{-+}$.
In Fig.~\ref{fig:chipm_dimers}(a) we show the imaginary part of the local $\chi_{-+}$ component as a function of the frequency for the nearest neighbor dimer (blue curve).
The optical mode is out of the range of the figure.
As before, the small peak at negative frequencies reveals an elliptical precession.
The properties of the precessional motion can be gleaned from Figs.~\ref{fig:chipm_dimers}(b) and (c). 
For the peak at positive frequencies, the amplitude is large, while the movement is slightly elliptical ($e\sim0.5$). 
On the other hand, for the negative resonance, the amplitude is small but significant, while the precession becomes linear for frequencies close to $\omega^-$.
The low value of the damping obtained for this dimer leads to a relatively sharp peak in the blue curve of Fig.~\ref{fig:chipm_dimers}(a).
This is due to the strong $d$-$d$ hybridization, which creates bonding and antibonding states, thus lowering the density of states near the Fermi energy.
As the damping parameter is very sensitive to the electronic structure around the Fermi energy \cite{Lounis:2015ho}, this explains its reduction to about a third of the values for the Fe adatoms.

In the following section, we uncover the dynamical behavior when the interatomic exchange coupling is now of the same order of magnitude as the magnetic anisotropy constants.
The two spin moments are no longer forced to be parallel to each other, and this has important consequences for the precessional motion.

\subsection{Dimer II}

\begin{figure}[!htb]
  \centering
   \includegraphics[width=0.5\columnwidth]{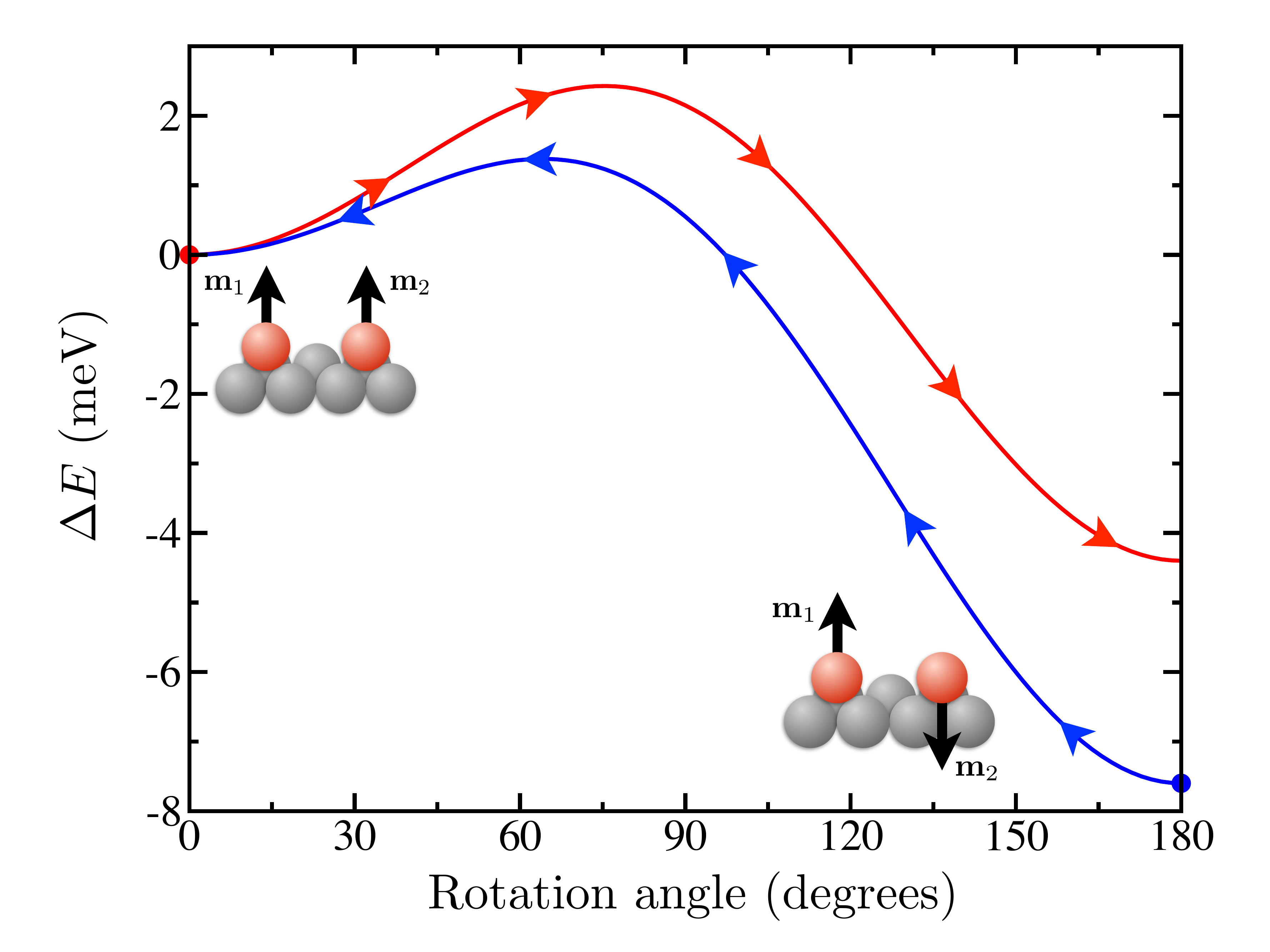}
 \caption{
    Stability of the two possible magnetic alignments for dimer II from the variation of the band energy.
The energy is plotted as a function of the angle between the magnetic moments of the impurities with $\mathbf{M}_1$ fixed to $\VEC{\hat{n}}_{z'}$ while $\mathbf{M}_2$ is rotated in the $x'z'$ plane. We take as a reference the ferromagnetic state, i.e., $\Delta E = E^\text{band}(M_1\VEC{\hat{n}}_{z'},\mathbf{M}_2) - E^\text{band}(M_1\VEC{\hat{n}}_{z'},M_2\VEC{\hat{n}}_{z'})$.
Red and blue curves are obtained starting from a self-consistent ferromagnetic and antiferromagnetic states, respectively. 
}
 \label{fig:FeFe_dimer_2nn_barrier}
\end{figure}

When the two Fe atoms are pulled apart, their properties quickly recover those of isolated units.
For a separation of twice the nearest neighbor distance, the spin and orbital moments are very close to the ones obtained for the single Fe adatom, as seen in Table~\ref{tab:parameters_dimer}.
Their coupling is weakened and even changes sign, as indicated by the value obtained from the total energy difference $J^{\Delta E}$ and from the magnetic force theorem $J_0$.
As these values are lower than the anisotropy constants of the isolated adatoms, the interplay between these two kinds of magnetic interactions may give rise to a metastable state.
This is confirmed by Fig.~\ref{fig:FeFe_dimer_2nn_barrier}, where we plot the band energy variations starting from two self-consistent states, FM (red curve) and AFM (blue curve), as a function of the angle between the magnetic moments.
We see that the AFM alignment is the ground state, but also that the FM alignment is a local minimum of the energy.

\begin{figure}[!htb]
  \centering
   \includegraphics[width=0.45\columnwidth]{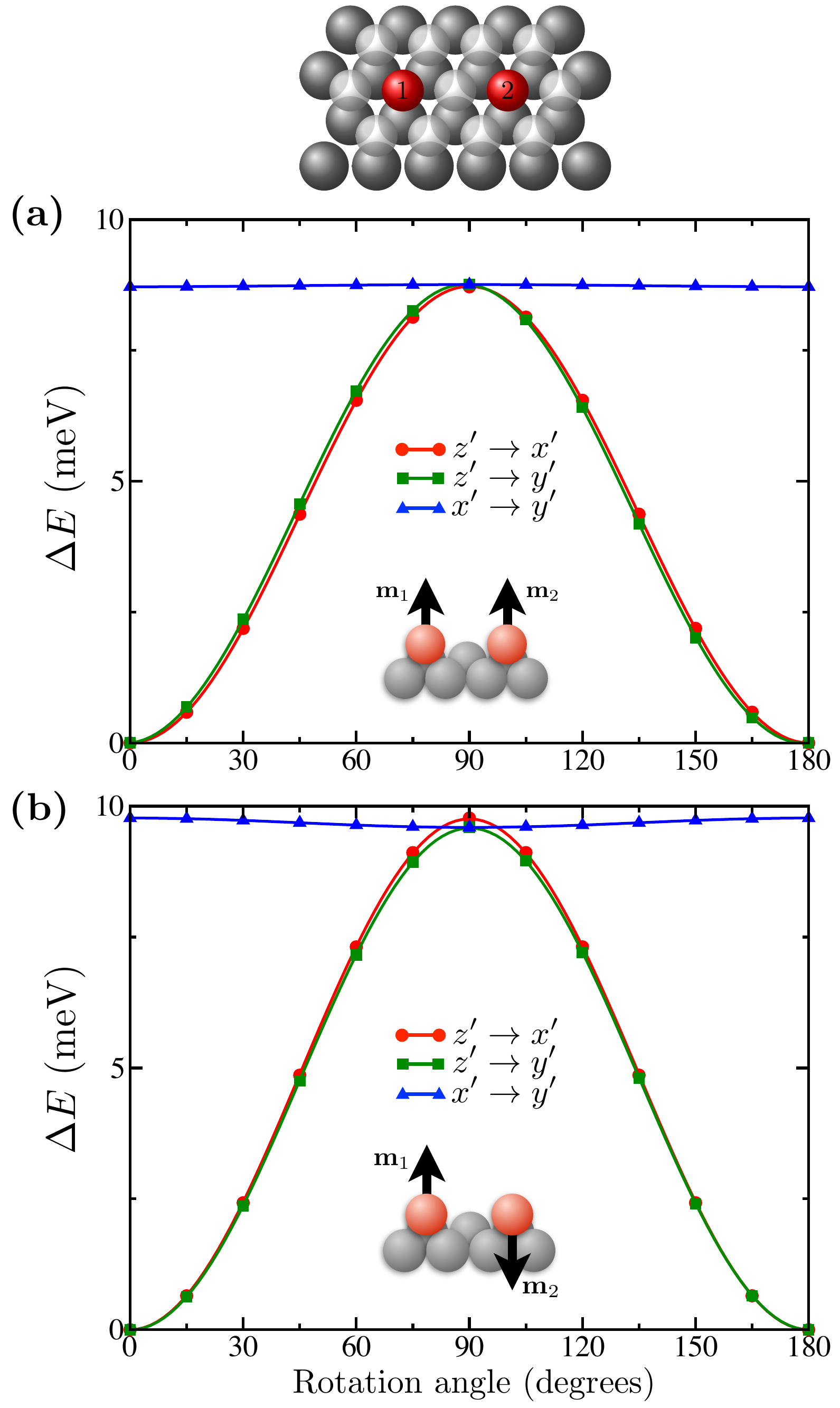}
 \caption{
   Variation of the band energy for magnetic anisotropy energy of the dimer II structure.
(a) The energy is plotted as a function of the magnetization direction of the parallel alignment (as in the central diagram), $\mathbf{M} = (M_1+M_2)\VEC{\hat{n}}$, taking as a reference the self-consistent calculation where $\mathbf{M}\|\VEC{\hat{n}}_{z'}$, i.e., $\Delta E = E^\text{band}(\mathbf{M}) - E^\text{band}(M\VEC{\hat{n}}_{z'})$.
(b) Change of the band energy as a function of the magnetization direction of the antiparallel alignment (as in the central diagram), $\mathbf{M} = (M_1-M_2)\VEC{\hat{n}}$, taking as a reference the self-consistent calculation where $\mathbf{M}\|\VEC{\hat{n}}_{z'}$, i.e., $\Delta E = E^\text{band}(\mathbf{M}) - E^\text{band}(M\VEC{\hat{n}}_{z'})$.
The dimer is composed of two Fe atoms (1 and 2) situated at twice the nearest neighbor distance on the Cu(111) surface, as illustrated in the central diagram.
Dark gray spheres represent Cu atoms and transparent spheres are vacuum positions around the Fe sites. 
The anisotropy constants are obtained by fitting the energy curves when the magnetization of both sites are rotated simultaneously along the $z'x'$ plane (red circles), the $z'y'$ plane (green squares), and along the $x'y'$ plane (blue triangles).
The values of the magnetic anisotropy components in the local frame of reference are listed in Table~\ref{tab:parameters_dimer}.}\label{fig:FeFe_mae_2nn}
\end{figure}

As both magnetic states are accessible, we characterize their magnetic anisotropic energy as done before for dimer I.
The variation of the energy when the moments are rotated together in the FM or the AFM alignments, shown in Fig.~\ref{fig:FeFe_mae_2nn}, indicates an almost uniaxial magnetic anisotropy energy, with $K_x\simeq K_y$.
The obtained values are given in Table~\ref{tab:parameters_dimer}.
The anisotropy constants are essentially independent from the magnetic alignment and are very close to the values found for the isolated adatom, which shows that the Fe adatoms weakly disturb each other.

We can access two qualitatively different kinds of spin excitations, by starting either from the AFM ground state or from the FM metastable state.
To compare with dimer I, we consider again the local susceptibility $\chi_{1-,1+}\equiv\chi_{-+}$ and we first analyze the FM case.
The corresponding density of spin excitations is the red curve of Fig.~\ref{fig:chipm_dimers}(a).
As before, we have the two modes, acoustic and optical, described by Eq.~\eqref{eigenenergies_dimer_fm_alpha}.
Since $J<0$, the optical mode has lower energy than the acoustic mode: While the latter is located at $\sim\SI{5}{\milli\electronvolt}$, the former resonates at $\sim\SI{2}{\milli\electronvolt}$.
Note that, unlike the biaxial case of dimer I, no signal can be seen for negative energies in the inset of Fig.~\ref{fig:chipm_dimers}(a), indicating a nearly circular precession.
The amplitude of the motion for the optical mode is larger than for the acoustic, see Fig.~\ref{fig:chipm_dimers}(b).
The eccentricity of precession, represented by the red curve of Fig.~\ref{fig:chipm_dimers}(c), reaches a maximum value of $e\sim0.5$ close to $\omega^-_\text{op}$.
This confirms that the precession is slightly elliptical.

A completely different excitation spectrum is obtained when the initial state is, instead, the AFM ground state, as seen in the green curve of Fig.~\ref{fig:chipm_dimers}(a).
Note that even for the almost uniaxial anisotropy case considered here, the transverse susceptibility presents a peak at negative energies.
However, Fig.~\ref{fig:chipm_dimers}(b) shows that the amplitude of precession is significant only for the peak at positive frequencies and is featureless for negative frequencies.
In addition, the precessional motion is only slightly elliptical ($e<0.5$), see Fig.~\ref{fig:chipm_dimers}(c).

We can interpret this behavior as reflecting the different preferred precessional senses of the two Fe atoms composing the dimer. 
If the two atoms were uncoupled ($J=0$), the intrinsic precessional sense of atom 1 would be counterclockwise, while atom 2 (being antiparallel to atom 1) would naturally precess clockwise, in a common frame of reference.
We assume that the tunneling current is exciting the precessional motion of atom 1.
For positive frequencies, the excitation spectrum of atom 1 is similar to that of the isolated adatom, cf. Fig.~\ref{fig:chipm_adatom}(a), which indicates that the coupling $J$ to atom 2 is not playing a significant role.
For negative frequencies, if atom 1 was isolated, we would not expect any resonant behavior. 
However, the unfavorable precession driven in atom 1 is transferred to atom 2 via $J$, triggering its natural precessional motion, which then feeds back to atom 1 (again via $J$) leading to the observed enhanced response.
In fact, the presence of the excitation at negative energies indicates that the classical antiferromagnetic state $\ket{\uparrow\downarrow}$, for which we obtain the spectra, is not the true ground state of the system \cite{Anderson:1952jp}.


As done for dimer I, we can obtain analytical expressions for the resonances considering biaxial anisotropy and damping:
\begin{equation}\label{eigenenergies_dimer_afm_biaxial_alpha}
\begin{split}
\frac{\omega^\pm_1}{\gamma'}=&-\iu\alpha(K_x+K_y-J)\\
&\pm\sqrt{4K_x(K_y-J)-\alpha^2(J+K_x-K_y)^2}\\
\frac{\omega^\pm_2}{\gamma'}=&-\iu\alpha(K_x+K_y-J)\\
&\pm\sqrt{4K_y(K_x-J)-\alpha^2(J-K_x+K_y)^2}
\end{split}\quad,
\end{equation}
where $\gamma' = \frac{\gamma}{M(1+\alpha^2)}$.
Here, $K_x,K_y>0$ and $J<0$ for the AFM ground state.
There are four solutions that correspond to two distinct precessional modes (the $+$ and $-$ solutions generate the same precessional motion, as explained in Appendix~\ref{apx:ellipticalmode}).
If the biaxial character is very weak, $K_x\sim K_y$, the solutions become degenerate and we observe only two peaks in the spectrum instead of the expected four.
This is precisely the behavior that we obtain for dimer II.


The damping plays a much more important role for antiferromagnets than for ferromagnets, especially when the coupling is relatively large.
Increasing the damping strength, the resonance frequency lowers, as follows from Eq.~\eqref{eigenenergies_dimer_afm_biaxial_alpha}.
For simplicity, we explore this scenario for uniaxial anisotropy, $K_x=K_y=K$.
When $\alpha > 2\sqrt{K(K+|J|)}/|J|$, one of the modes moves to zero frequency, recovering the Goldstone mode, with the other one moving to the imaginary frequency axis, corresponding to an overdamped precession.
Although this condition is not fulfilled for our case, it may happen for dimers with the coupling $|J|\gg|K|$ (for example, a Mn dimer deposited on metallic substrates).

Finally, we consider the possible impact of the spin excitations on the stability of the classical AFM ground state.
In Ref.~\onlinecite{Holzberger:2013du}, it was argued that the AFM dimer was able to access the two degenerate N\'eel states $\ket{\uparrow\downarrow}$ and $\ket{\downarrow\uparrow}$, since the zero-point fluctuations (involving the coupling $J$) were larger than the energy barrier between them (proportional to $K$).
In that work, the energy of the zero-point fluctuations was connected to the energy of the lowest spin excitation mode.
This description is also in accordance with Ref.~\onlinecite{Anderson:1952jp}, where the zero-point energy was found to vanish for an AFM dimer with zero anisotropy.
Nevertheless, neither of these works have considered the effects of damping.
Our results obtained in Eq.~\eqref{eigenenergies_dimer_afm_biaxial_alpha} show that the damping reduces the frequency of the lowest excitation mode for antiferromagnetic dimers with large coupling.
Counterintuitively, this implies that the zero-point energy is also lowered when the damping increases, following the argument of Ref.~\onlinecite{Holzberger:2013du}.
This may then prevent the fluctuations over the energy barrier between the two N\'eel states and stabilize them.
This can be contrasted with the behavior found in Ref.~\onlinecite{IbanezAzpiroz:2016fa}, in which the damping increases the zero-point spin fluctuations of single magnetic adatoms deposited on metallic surfaces.


\section{Conclusions}
\label{sec:Summary}

In this paper, we have presented a semiclassical interpretation of the dynamical spin excitations in magnetic nanostructures computed using a first-principles approach.
A crucial role is played by the spin-orbit coupling, responsible for nontrivial magnetic anisotropies, which in turn lead to complex precessional motion.
A description of the general elliptical precession was provided and connected to the spectral features of the transverse magnetic susceptibility.
As the latter is intimately related to the inelastic contribution to the tunneling conductance, we believe this formalism can provide useful insights on the nature and properties of spin excitations detected experimentally.

Considering a single Fe adatom deposited on the Cu(111) surface, we showed how the ground state and also the excitation properties can be controlled by an external magnetic field or by atomic manipulation of its environment (the formation of a Fe-Cu dimer). 
We found that the signature of noncircular precessional motion is the appearance of a secondary peak at negative frequencies in the density of spin excitations.
In the vicinity of this peak, the precession becomes highly elliptical.
Our results indicate that the spin pumping mechanism is quite isotropic, while the precessional damping is large, anisotropic and tunable.

We next considered a different kind of atomic manipulation, where the Cu atom is replaced by a second Fe atom.
When the two Fe atoms are nearest neighbors, they are strongly ferromagnetically coupled, behaving as a single magnetic unit when the frequency is much lower than their coupling.
The magnetic anisotropy is now of biaxial nature, leading to elliptical spin excitations.
The dynamical properties can be understood from the strong $d$-$d$ hybridization, which is responsible for lowering both the spin pumping efficiency and the magnetic damping.

When the Fe adatoms are pulled apart, their coupling weakens and becomes antiferromagnetic, being now comparable in magnitude to the magnetic anisotropy energy.
This special configuration gives access to two different magnetic states: the antiferromagnetic ground state and the metastable ferromagnetic state.
The two Fe atoms weakly influence each other, and their local properties are very similar to those of isolated Fe adatoms.
The metastable FM state leads to two kinds of spin excitations: an acoustic mode where the spins precess in-phase and an optical mode where they precess in anti-phase.
The optical mode actually has lower energy than the acoustic one, as a consequence of the metastable nature of the ferromagnetic alignment.
By inverting the spin alignment, we arrive at the antiferromagnetic ground state, which has completely different excitation characteristics.
We find only one broad excitation peak at positive energies, instead of the two for the ferromagnetic alignment.
The antiparallel Fe atoms have opposite intrinsic precessional motion, which leads to a secondary peak at negative energy.
This does not represent elliptical precession, in contrast to the strongly ferromagnetic dimer.
We obtain the excitation energies of the system, and show that the damping contribution lowers the resonance frequency, specially for antiferromagnetic dimers with large coupling.
The lowest energy mode is connected to the zero-point fluctuation energy, which indicates that this lowering may inhibit fluctuations over the barrier between different N\'eel states.

Our results shed light on how the spin excitations can be engineered by bringing atoms together or separating them apart.
We demonstrate how the external fields, magnetic anisotropy energies, exchange couplings and damping, as well as the initial alignment between the magnetic units can be used to design a diverse range of precessional motions.
Two particular outcomes may be used as experimental guidance.
First, we propose a pump-probe-like experiment \cite{Loth:2010gq} to access the influence of the magnetic alignment on the dynamical spin excitations, while keeping all the other quantities essentially unchanged: The system is put into the metastable state by an initial perturbation (pump) followed by a measurement of its excitations by a probe.
They can be compared to the excitations from the ground state, when the pump is switched off.
Second, the effect of the damping on the zero-point fluctuation energy indicate that STM experiments made on dimers with similar magnetic interactions but different damping strengths (e.g., deposited on insulators or metals) may present rather different magnetic signals.
The effect of the damping on the excitation modes also impacts the field of antiferromagnetic spintronics.
A naive expectation is that ultrafast antiferromagnetic devices shall involve large coupling between the units, to make them switch together, and high damping, to quickly relax the magnetization to a new switched state.
Our results demonstrate that the correct picture is more subtle, and a combination of anisotropy, coupling and damping must be taken into account.

The interplay between the different magnetic interactions offers multiple tools to control processing speeds and polarization of magnetic units and emitted spin currents, which may lay the foundations of the building blocks of future devices.
Our first-principles description of the dynamical properties of magnetic nanostructures provides a predictive approach to the design and engineering of those building blocks.

\begin{acknowledgments}

This work is supported by the European Research Council (ERC) under the European Union's Horizon 2020 research and innovation programme (ERC-consolidator Grant No. 681405 -- DYNASORE).

\end{acknowledgments}

\begin{appendix}

\section{Phenomenological expressions for the dynamical susceptibiltiies of nanostructures}
\label{apx:eqsmotion}

By linearizing the LLG equation of motion [Eq.~\eqref{llg}], they can be written in the local frame of reference as $[\chi^{-1}(\omega)] \delta\mathbf{M} = \delta\mathbf{B}^{\text{ext}}$.
In this appendix, we list the inverse susceptibilities $\chi^{-1}$ used to fit the TDDFT results and obtain the parameters of the phenomenological model.
In the following, we use $\gamma_\pm = \frac{\gamma}{1\pm \iu \alpha}$.

\subsection{Single atom}

For adatoms, $\delta\mathbf{M}$ and $\delta\mathbf{B}^\text{ext}$ are vectors containing the transverse circular components $+,-$ of the magnetization and the external field, respectively.

\subsubsection{Uniaxial, no magnetic field}
\label{adatom_uniaxial}

When the magnetization points perpendicularly to the surface, the symmetry of the system is $C_{3v}$ and the anisotropy is uniaxial --- in our convention, $K_x = K_y = K$ and $K_z=0$. The inverse susceptibility for this case is given by

\begin{equation}\label{adatom_uniaxial+-}
\begin{split}
\chi^{-1}(\omega) = \left(\begin{array}{cc} 
\frac{2K}{M^2} +\frac{\omega}{\gamma_- M} & 0\\
0 & \frac{2K}{M^2}-\frac{\omega}{\gamma_+ M} 
\end{array}\right)
\end{split}\quad.
\end{equation}

\subsubsection{Uniaxial, magnetic field along $\VEC{\hat{n}}_{y'}$}
\label{adatom_uniaxial_bext}

The in-plane magnetic field saturates the magnetization along the $\VEC{\hat{n}}_{y'}$ direction in the global frame of reference. Transforming to the local frame of reference, where it points along $\VEC{\hat{n}}_z$, we find $K_x = K_z = 0$ and $K_y=-K$.
The inverse susceptibility is

\begin{equation}\label{adatom_uniaxial_bext+-}
\begin{split}
\chi^{-1}(\omega) =\left(\begin{array}{cc} 
\frac{MB^\text{ext}_0-K}{M^2} +\frac{\omega}{\gamma_- M} & \frac{K}{M^2} \\
\frac{K}{M^2} & \frac{MB^\text{ext}_0-K}{M^2}-\frac{\omega}{\gamma_+ M}
\end{array}\right)
\end{split}\quad.
\end{equation}

\subsubsection{Biaxial}
\label{adatom_biaxial}

Placing a Cu atom close to the Fe adatom, we break the symmetry and change the magnetic anisotropy landscape of the system. 
By a suitable choice of the local frame of reference, the anisotropy matrix can be written in the diagonal form given in Eq.~\eqref{kmatrix}.
$\chi^{-1}$ is then

\begin{equation}\label{atadom_biaxial+-}
\begin{split}
\chi^{-1}(\omega) = \left(\begin{array}{cc} 
\frac{K_x+K_y}{M^2} +\frac{\omega}{\gamma_- M} & \frac{K_x-K_y}{M^2}\\
\frac{K_x-K_y}{M^2} & \frac{K_x+K_y}{M^2}-\frac{\omega}{\gamma_+ M}
\end{array}\right)
\end{split}\quad.
\end{equation}

\subsection{Dimer}

Dimer structures naturally break the symmetry of the system. 
The linearized equation of motion includes the transverse components of both magnetic units in the global frame of reference, $\delta\mathbf{M} = (M_{1+},M_{2+},M_{1-},M_{2-})$.

\begin{widetext}

\subsubsection{Parallel alignment}
\label{dimer_ferro}

When the moments are pointing parallel to each other, the inverse susceptibility is
\begin{equation}\label{dimer_ferro+-}
\begin{split}
\chi^{-1}(\omega) = \left(\begin{array}{cccc} 
\frac{(K_x+K_y)+J}{M^2} +\frac{\omega}{\gamma_- M} & - \frac{J}{M^2} & \frac{(K_x-K_y)}{M^2} & 0\\
- \frac{J}{M^2} & \frac{(K_x+K_y)+J}{M^2}+\frac{\omega}{\gamma_- M} & 0 & \frac{(K_x-K_y)}{M^2} \\
\frac{(K_x-K_y)}{M^2} & 0 & \frac{(K_x+K_y)+J}{M^2}-\frac{\omega}{\gamma_+ M} & - \frac{J}{M^2} \\
0 & \frac{(K_x-K_y)}{M^2} & - \frac{J}{M^2} & \frac{(K_x+K_y)+J}{M^2}-\frac{\omega}{\gamma_+ M}
\end{array}\right)
\end{split}
\end{equation}

\subsubsection{Antiparallel alignment}
\label{dimer_anti}

In the case of antiparallel alignment of the magnetic moments, we have
\begin{equation}\label{dimer_anti+-}
\begin{split}
\chi^{-1}(\omega) = \left(\begin{array}{cccc}
 \frac{(K_x+K_y)-J}{M^2}+\frac{\omega}{\gamma_- M} & \frac{J}{M^2} & \frac{(K_x-K_y)}{M^2} & 0 \\ 
\frac{J}{M^2} & \frac{(K_x+K_y) -J}{M^2}-\frac{\omega}{\gamma_+ M} & 0 & \frac{(K_x-K_y)}{M^2} \\
\frac{(K_x-K_y)}{M^2} & 0 & \frac{(K_x+K_y)-J}{M^2}-\frac{\omega}{\gamma_+ M} & \frac{J}{M^2} \\
0 & \frac{(K_x-K_y)}{M^2} & \frac{J}{M^2} & \frac{(K_x+K_y) -J}{M^2}+\frac{\omega}{\gamma_- M}
\end{array}\right)
\end{split}\quad.
\end{equation}

\end{widetext}

\section{Elliptical mode of single atoms in a circular basis}
\label{apx:ellipticalmode}

To obtain the natural precessional modes of a single adatom, we linearize the time-dependence of the magnetic moment (see Sec.~\ref{sec:ellipse}).
The small transverse components can be found from the associated LLG equation $\chi^{-1}\delta\mathbf{M} = \delta \mathbf{B}^\text{ext}$.
The normal modes of precession are obtained by solving the secular problem $\delta \mathbf{B}^\text{ext}=0$, which leads to the eigenvalue problem $D| u\rangle = \omega |u\rangle$, or 
\begin{equation}
\begin{split}
\left(\begin{array}{cc} \gamma_+ B_\| & \gamma_+ B_\perp \\ -\gamma_- B_\perp^* & -\gamma_- B_\|^* \end{array}\right)\left(\begin{array}{c}u_1  \\ u_2\end{array}\right)= \omega \left(\begin{array}{c}u_1  \\ u_2\end{array}\right)
\end{split}\quad .
\end{equation}
$D$ is the dynamical matrix written in the circular basis with eigenmodes given by the poles of the susceptibility, obtained in Eq.~\eqref{eigenenergies}. Since $\omega^- = -(\omega^+)^*$, the eigenvectors are obtained from
\begin{equation}
\begin{split}
|u^+\rangle:& \quad (\gamma_+ B_\| -\omega^+) u^+_1 + \gamma_+ B_\perp u^+_2 = 0\\
|u^-\rangle: &  \quad (\gamma_+ B_\| -\omega^+) u^{-*}_{2} + \gamma_+ B_\perp u^{-*}_{1} = 0
\end{split}\quad,
\end{equation}
which shows that $u^{-*}_{2} = u^{+*}_1$ and $u^{-*}_{1} = u^{+}_2$. Therefore, we can write for the normalized eigenvectors
\begin{equation}
\begin{split}
|u^+\rangle = \left(\begin{array}{c} e^{\iu(\phi_0-\phi)} \cos\xi \\ e^{\iu(\phi_0+\phi)} \sin\xi \end{array}\right) \, , \,
|u^-\rangle = \left(\begin{array}{c} e^{-\iu(\phi_0+\phi)} \sin\xi \\ e^{-\iu(\phi_0-\phi)} \cos\xi \end{array}\right)
\end{split}\quad,
\end{equation}
where we have defined
\begin{equation}
\begin{split}
e^{\iu(\phi_0-\phi)} \cos\xi = \frac{-\gamma_+ B_\perp}{\sqrt{|\gamma_+ B_\perp|^2+|\gamma_+ B_\| -\omega^+|^2}}\\
e^{\iu(\phi_0+\phi)} \sin\xi = \frac{\gamma_+ B_\| -\omega^+}{\sqrt{|\gamma_+ B_\perp|^2+|\gamma_+ B_\| -\omega^+|^2}}
\end{split}\quad.
\end{equation}

Notice that, since the dynamical matrix $D$ is not Hermitian, the eigenvectors $|u^+\rangle$ and $|u^-\rangle$ are not orthogonal to each other but rather to the left eigenvectors. These may be obtained by $\langle v | D = \omega \langle v |$. 
\begin{equation}
\begin{split}
\langle v^+| =& \left(\begin{array}{cc} e^{-\iu(\phi_0-\phi)} \cos\xi & -e^{-\iu(\phi_0+\phi)} \sin\xi \end{array}\right) \\
\langle v^-| =& \left(\begin{array}{cc} -e^{\iu(\phi_0+\phi)} \sin\xi & e^{\iu(\phi_0-\phi)} \cos\xi \end{array}\right)
\end{split}\quad.
\end{equation}
The left and right eigenvectors are orthogonal, i.e., $\langle v^+| u^-\rangle = \langle v^-| u^+\rangle = 0$, and $\langle v^+| u^+\rangle = \langle v^-| u^-\rangle = \cos^2\xi-\sin^2\xi$. The dynamical matrix can then be written as
\begin{equation}
\begin{split}
D = \omega^+ \frac{| u^+\rangle\langle v^+|}{\langle v^+| u^+\rangle}+\omega^- \frac{| u^-\rangle\langle v^-|}{\langle v^-| u^-\rangle}
\end{split}\quad.
\end{equation}

Since $D$ is written in the circular basis, the eigenvectors represent the transverse components of the magnetization $\delta M_-$ and $\delta M_+$.
Both $|u^+\rangle$ and $|u^-\rangle$ lead to the same time dependence of the magnetization.
Making use of Eq.~\eqref{changebasis}, we finally find
\begin{equation}\label{transvmag}
\begin{split}
\delta\mathbf{M}(t) = &\frac{M\delta\theta}{2} \operatorname{Re}\left\{\left[ \left(e^{-\iu\phi}\cos\xi+e^{\iu\phi}\sin\xi\right)\VEC{\hat{n}}_x\right.\right.\\
&\left.\left.+\iu\left(e^{-\iu\phi}\cos\xi-e^{\iu\phi}\sin\xi\right)\VEC{\hat{n}}_y\right] e^{\iu(\phi_0-\omega' t)}\right\}e^{\omega'' t}
\end{split}\quad.
\end{equation}
where we have used $\omega^\pm = \pm\omega'+\iu\omega''$, with $\omega'$ and $\omega''$ the real and imaginary part of the frequency ($\omega''<0$), respectively.

In general, the modes are elliptical, as both $\delta M_+$ and $\delta M_-$ are finite.
When applying time-dependent magnetic fields, the susceptibility can also be written in terms of the normal modes as
\begin{equation}
\begin{split}
&\left(\begin{array}{c}\delta M_- \\ \delta M_+\end{array}\right)=\\
 &\quad\left[\frac{M}{\omega^+-\omega} \frac{| u^+\rangle\langle v^+|}{\langle v^+| u^+\rangle}+\frac{M}{\omega^--\omega} \frac{| u^-\rangle\langle v^-|}{\langle v^-| u^-\rangle}\right] \left(\begin{array}{c}\delta B_-^\text{ext} \\ \delta B_+^\text{ext}\end{array}\right)
\end{split}\quad.
\end{equation}
A counterclockwise circularly polarized external field $\delta B_0^\text{ext} \left(\begin{array}{c} 1 \\ 0 \end{array}\right)$ is not orthogonal to either $\langle v^+|$ or $\langle v^-|$ in general, which leads to peaks in the transverse dynamical susceptibility $\chi_{-+}(\omega)$ at both $\omega^+$ and $\omega^-$ (as discussed in Secs.~\ref{sec:adatom} and \ref{sec:dimer}).

\end{appendix}

\bibliography{dimer.bib}
\end{document}